\documentclass[pdflatex,sn-nature]{sn-jnl}
\PassOptionsToPackage{bookmarksopenlevel=10}{hyperref}

\usepackage{graphicx}%
\usepackage{multirow}%
\usepackage{amsmath,amssymb,amsfonts}%
\usepackage{amsthm}%
\usepackage{mathrsfs}%
\usepackage[title]{appendix}%
\usepackage{xcolor}%
\usepackage{textcomp}%
\usepackage{manyfoot}%
\usepackage{booktabs}%
\usepackage{algorithm}%
\usepackage{algorithmicx}%
\usepackage{algpseudocode}%
\usepackage{listings}%
\usepackage[per-mode=symbol]{siunitx}%
\usepackage{bm}%
\usepackage{subcaption}
\usepackage[utf8]{inputenc}
\usepackage{verbatim}
\usepackage{enumitem}

\theoremstyle{thmstyleone}%
\theoremstyle{thmstyletwo}%

\theoremstyle{thmstylethree}%
\newcommand{\mrm}{\mathrm}
\newcommand{\Heff}{H_\mathrm{eff}}
\newcommand{\Hext}{H_\mathrm{ext}}
\newcommand{\PdB}{\mathrm{P}_\mathrm{1dB}}
\newcommand{\BWthreeDB}{\mathrm{BW}_\mathrm{3dB}}
\newcommand{\dm}{d_\mathrm{m}}
\DeclareSIUnit\dBm{dBm}

\begin{document}

\title[Article Title]{Microscaled Tunable Magnonic RF Phase Shifters}


\author*[1]{\fnm{Johannes} \sur{Greil}}\email{johannes.greil@tum.de}
\author[2]{\fnm{Antonio} \sur{Angotti}}
\author[3]{\fnm{Felix} \sur{Kohl}}
\author[4]{\fnm{\'{A}d\'{a}m} \sur{Papp}}
\author[3]{\fnm{Matthias} \sur{Wagner}}
\author[2]{\fnm{Maria} \sur{Cocconcelli}}
\author[2]{\fnm{Andrea} \sur{Del Giacco}}
\author[5]{\fnm{Dieter} \sur{Ferling}}
\author[3]{\fnm{Bj\"{o}rn} \sur{Heinz}}
\author[2]{\fnm{Federico} \sur{Maspero}}
\author[4]{\fnm{Gy\"{o}rgy} \sur{Csaba}}
\author[2]{\fnm{Riccardo} \sur{Bertacco}}
\author[1]{\fnm{Markus} \sur{Becherer}}
\author*[3]{\fnm{Philipp} \sur{Pirro}}\email{ppirro@rptu.de}

\affil[1]{\orgdiv{Professorship of Chip-Based Magnetic Sensor Technology}, \orgname{Technical University of Munich}, \country{Germany}}

\affil[2]{\orgdiv{Department of Physics and Polifab}, \orgname{Politecnico di Milano}, \country{Italy}}

\affil[3]{\orgdiv{Fachbereich Physik and Landesforschungszentrum OPTIMAS}, \orgname{Rheinland-Pf\"{a}lzische Technische Universit\"{a}t Kaiserslautern-Landau}, \country{Germany}}

\affil[4]{\orgdiv{Faculty of Information Technology and Bionics}, \orgname{P\'{a}zm\'{a}ny P\'{e}ter Catholic University}, \city{Budapest}, \country{Hungary}}

\affil[5]{\orgdiv{Nokia Bell Labs}, \orgname{Nokia Corporation}, \city{Stuttgart}, \country{Germany}}


\abstract{Tunable, microscopic, and energy-efficient solutions for radio-frequency (RF) signal manipulation in the GHz regime are a key technology for efficient communication and sensing applications. Spin waves offer micrometer wavelengths at GHz frequencies, combined with strong magnetic-field tunability, making them inherently well-suited for tunable analog signal processing. Here, we demonstrate a novel concept: a micron-scale tunable RF phase shifter based on the wavelength shift of propagating spin waves. High energy efficiency is achieved by using the stray field of a micromagnet on a piezoelectrically actuated MEMS cantilever to locally induce this shift. The device shows a phase shift of more than \SI{360}{\degree} at a center frequency of \SI{6.1}{\giga\hertz} using a phase-shifting area of less than \SI{0.02}{\milli\meter}$^2$ and a channel footprint of \SI{0.5}{\milli\meter}$^2$. By changing the magnetic bias field, its functionality is experimentally confirmed over a range of center frequencies from \SI{3}{\giga\hertz} to \SI{8.2}{\giga\hertz}, and simulations show its applicability up to \SI{14}{\giga\hertz}. A system-level characterization of an embedded device version demonstrates the qualification of magnonic phase shifters for highly integrated RF systems.}


\maketitle

\section{Introduction}
\label{sec:Introduction}

Today’s information-driven society relies on electromagnetic radio-frequency (RF) waves in the Gigahertz (GHz) regime for wireless communication, radar, sensing, and satellite technologies. 
The increasing demand for higher data throughput continuously pushes RF front-end systems toward broader and higher-frequency bands, creating a growing need for compact, tunable, and highly integrated RF components~\cite{Qin_Li_Liu_Liao_Li_Qiu_2024,Nguyen_2005}. 
Among these, tunable phase shifters are key building blocks in phased-array systems~\cite{Wu_Mehlman_Kumar_Moy_Jia_Ma_Wagner_Sturm_Verma_2021}, where they enable electronic beam steering and beam forming without mechanical antenna movement. 
Such systems are essential for radar~\cite{Stailey_Hondl_2016}, wireless communication~\cite{Naqvi_Lim_2018}, and satellite links~\cite{Kebe_Yagoub_Amaya_2025}. 
For the performance of RF phase shifters, a central challenge is to tune the phase of an RF signal in a wide range while maintaining stable transmission characteristics, including insertion loss (IL), return loss (RL), center frequency ($f_\mrm{c}$), and bandwidth (BW).
For integrated systems, the power consumption and channel area are additional key design parameters~\cite{Kebe_Yagoub_Amaya_2025}.

Existing RF phase shifters employ a variety of physical mechanisms, including electronic, magnetic, mechanical, and micro-electromechanical systems (MEMS)~\cite {Kebe_Yagoub_Amaya_2025}. 
Commercial analog phase shifters are often based on varactor diodes~\cite{pasternak_pe82p2000,pasternak_pe82p2001,Qorvo_CMD297,MiniCircuits_SMD_JSPHS-661}, while MEMS-based implementations use tunable resonators or switched transmission-line segments~\cite{Rebeiz_MEMSphaseShifter_2002}. 
Alternative concepts exploit ferrites, liquid crystals, or ferroelectric structures~\cite{Cheng_PhaseShifter_2014,Sharma_PhaseShifter_2019,ALCAN_LC_PhaseShifter,Li_LCphaseShifter_2019,Acikel_PhaseShifter_2002}. 
Although these technologies provide different trade-offs between tuning range, bandwidth, and losses, achieving simultaneously low power consumption, compact footprint, broadband operation, and low insertion loss remains challenging for integrated RF systems.

Spin waves (SWs), the collective excitations of electron spins in magnetic materials, offer a promising alternative for miniaturized and energy-efficient RF signal processing. 
Due to their wavelengths ranging from tens of nanometers to tens of micrometers, SWs operate at three to four orders of magnitude shorter wavelengths than microwave photons at the same frequency, enabling compact analog phase manipulation. 
In addition, their operational point can be tuned via an external magnetic field, while the accessible wavelength range is determined by the microwave-to-SW transducer geometry, which can be designed to the needs of potential applications. 
These properties make SWs particularly attractive for reconfigurable, analog RF devices with small footprints and potentially low power consumption.

Magnetic-field-controlled microwave devices based on low-damping ferrimagnetic Yttrium-Iron-Garnet (YIG) have been investigated extensively, especially for tunable filter applications~\cite{Levchenko_Davidkova_Mikkelsen_Chumak_2026,Askarzadeh_YIGdevicesReview_2025,BritoBrito_TunableFilterReview_2011,Du_TunableBandpassFilter_2023,Yang_TunableBandpassFilter_2013,Shirkolaei_TunableFilter_2020,Kim_TunableFilter_2023}. 
Recent advances in material engineering and microfabrication have renewed interest in integrated magnonic RF devices with improved tuning ranges and insertion losses~\cite{Devitt_Wang_Tiwari_Bhave_2024,Devitt_2026_SW_filer_6G}. 
However, most demonstrated devices rely on localized resonances rather than propagating SWs for signal transmission. 
In contrast, propagating SWs enable large phase accumulation due to their low group velocity and micron-scale wavelengths at GHz frequencies, allowing the phase to be modified largely independently of the transmission amplitude. 
Previous concepts have demonstrated phase tuning using ferroelectric bilayers~\cite{Ustinov_Srinivasan_Kalinikos_2007,Ustinov_Kolkov_Nikitin_Kalinikos_Fetisov_Srinivasan_2011} or current-induced Oersted fields~\cite{Kostylev2007,Hansen_Demidov_Demokritov_2009}, but they require impractically high drive voltages for integrated applications, consume too much power, or have too large dimensions.

\section{Design of the Magnonic Phase Shifter}
\label{Sec:DeviceConcept}

\begin{figure}[ht!]
    \centering
    \includegraphics[width=\linewidth]{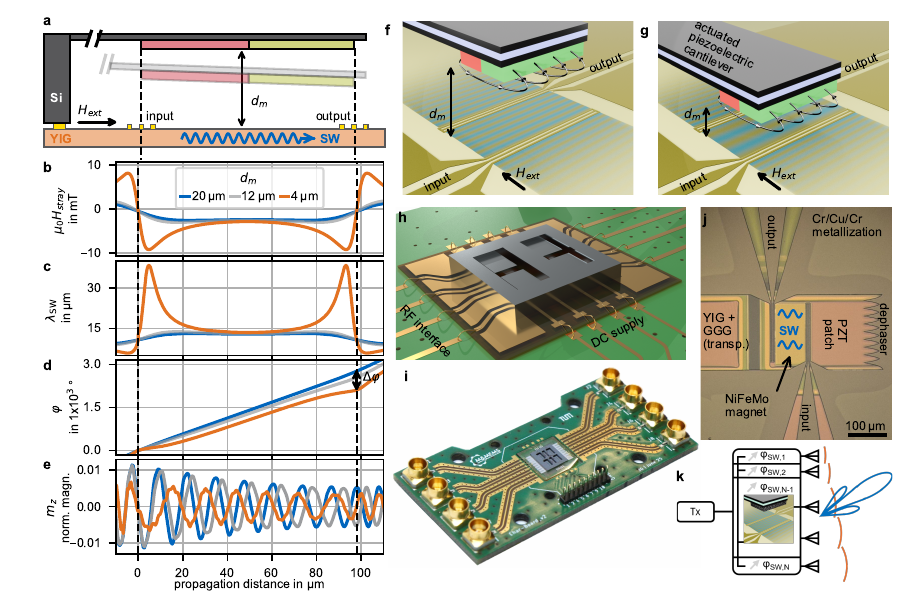}
    \caption{\textbf{Concept of a magnonic phase shifter with local magnetic field variation.} a) Schematic cross-section of the hybrid magnonic phase shifter consisting of a magnonic YIG chip and a silicon-based MEMS chip with piezoelectric cantilevers carrying micron-sized soft magnets. The cantilever actuation changes the magnet-to-chip distance $d_\mathrm{m}$, thereby locally tuning the effective magnetic field in the SW transmission channel. b) Analytically calculated stray-field profiles at the YIG surface for different $d_\mathrm{m}$ and ideal plan-parallel magnet actuation. c) Corresponding local SW wavelength variation and d) resulting accumulated phase along the channel. The graphs in b)-d) are extracted from an analytical model of the magnonic phase shifter's functionality. e) Micromagnetic simulations of the normalized magnetization showing the SW wavelength change for different cantilever positions, corresponding to phase shifts of approximately \SI{180}{\degree} and \SI{375}{\degree}. f)-g) The stray field is confined to the region between the transducers, allowing phase tuning without significantly altering the transmission characteristics. h) The hybrid chip wire-bonded to a PCB with RF and DC interfaces. i) Photograph of the assembled phase-shifter module as an example for RF integration. j) Through-substrate microscope image of the fabricated device showing the chip-to-chip alignment of about \SI{5}{\micro\meter}. k) Illustration of the integration of the magnonic phase shifter in phased-array RF systems for beam steering and beam forming.}
    \label{fig:Concept}
\end{figure}

Here, we present a voltage-tuned hybrid analog phase shifter with a phase tuning range $\Delta\varphi_\mathrm{max}$ exceeding \SI{375}{\degree} using a core phase-shifting area of approximately \SI{0.02}{\milli\meter}$^2$. 
The device exploits propagating SWs in the backward-volume geometry~\cite{Stancil2009} in a low-damping sub-micron-thick YIG film, where the phase accumulation is controlled by locally modifying the effective magnetic field $\Heff$ within the SW transmission channel. 
The phase shift originates from a local variation of the SW wave vector $k_\mrm{SW}(\Heff(x)) = 2\pi/\lambda_\mrm{SW}(\Heff(x))$ at a fixed frequency and propagation distance between the input and output microwave-to-SW transducers. 
The transducer geometry defines the excitable SW wavelength range and is designed around $\lambda_\mrm{SW}=\SI{10}{\micro\meter}$~\cite{Kohl_ModellingSW,Vanderveken_Lumped}. 
A schematic cross-section of the device is shown in Fig.~\ref{fig:Concept}a. 

The operating principle relies on two independently controlled static magnetic fields. 
A global external bias field $\Hext$ defines the operating point or center frequency $f_\mrm{c}$ according to $f_\mathrm{FMR} = \frac{\gamma \mu_0}{2\pi}\sqrt{\Hext(\Hext+M_\mrm{S})}$~\cite{Stancil2009} (gyromagnetic ratio $\gamma$, vacuum permeability $\mu_0$, saturation magnetization $M_\mrm{S}$), including a transducer-dependent offset from $f_\mathrm{FMR}$. 
A second, spatially localized field contribution is generated by the stray field $H_\mrm{stray}(x)$ of a micron-sized soft magnet, yielding $\Heff(x)=\Hext+H_\mrm{stray}(x)$. 
The soft magnet is integrated on a piezoelectric cantilever, enabling voltage-controlled tuning of $H_\mrm{stray}(x)$ through the magnet-to-chip distance $d_\mathrm{m}$. 
As SWs exhibit wavelengths orders of magnitude smaller than microwave photons at GHz frequencies, even small local changes of $\Heff$ strongly modify the accumulated phase along the propagation direction $x$. 
The resulting tunable phase accumulation is given by
\begin{equation}
\label{eq:phase_calc}
\varphi (d_\mathrm{m})= \int_{x_\mathrm{in}}^{x_\mathrm{out}}  k_\mrm{SW}(\Heff(x,d_\mathrm{m})) dx.
\end{equation}

A local variation of $H_\mrm{stray}(x)$ in the one-digit millitesla range (Fig.~\ref{fig:Concept}b) sufficiently modifies the SW wavelength (Fig.~\ref{fig:Concept}c) and therefore changes the accumulated phase $\varphi$ along the channel (Fig.~\ref{fig:Concept}d). 
To preserve the operating point of the SW transmission function, $H_\mrm{stray}(x)$ must remain spatially confined and symmetrically tuned within the SW channel. 
The phase shift is referenced to the largest magnet distance according to $\Delta\varphi=\varphi(d_\mathrm{m})-\varphi(d_\mathrm{m}=d_\mathrm{max})$. 
Precise alignment between the soft magnet and the SW channel is therefore essential for maximizing $\Delta\varphi$. 
Micromagnetic simulations of the propagating SWs are shown in Fig.~\ref{fig:Concept}e, where changing the cantilever position from $d_\mathrm{m}=\SIrange{20}{12}{\micro\meter}$ induces a phase shift of approximately \SI{180}{\degree}. 
Additional details of the phase calculation are provided in Supplementary Information Sec.~\ref{ssec:A_Model_Description_And_Usage}.

The hybrid device consists of two subsystems. 
The lower magnonic chip consists of a thin YIG film on a GGG substrate on which coplanar waveguide (CPW) transducers are fabricated that convert the RF signal into propagating SWs and back~\cite{Erdelyi_2025_Design_rules}. 
The upper silicon-based MEMS flip-chip integrates micron-sized soft magnets on piezoelectric cantilevers, as illustrated in Fig.~\ref{fig:Concept}f-g. 
Applying a DC voltage actuates the cantilevers and tunes the local stray field inside the SW transmission channel. 
The hybrid chip supports multiple SW channels and can be embedded and wire-bonded to a printed circuit board (PCB), as shown in Fig.~\ref{fig:Concept}h-i, providing RF and DC interfaces for system-level integration. 
A through-substrate microscope image of the assembled device is presented in Fig.~\ref{fig:Concept}j. 
The achieved chip-to-chip alignment accuracy of approximately \SI{5}{\micro\meter} is sufficient to position the soft magnets near the center of the \SI{98}{\micro\meter}-long and \SI{160}{\micro\meter}-wide SW channel. 
The resulting platform demonstrates the potential of hybrid magnonic phase shifters for compact RF systems such as phased arrays (Fig.~\ref{fig:Concept}k) used in beam steering and beam forming~\cite{Delos_PhasedArrays,Naqvi_Lim_2018,Wu_Mehlman_Kumar_Moy_Jia_Ma_Wagner_Sturm_Verma_2021}.

Three complementary models were used to design and optimize the phase shifter, as detailed in Supplementary Information, Sec.~\ref{ssec:A_Model_Description_And_Usage}. 
A semi-analytical phase shifter model calculates the accumulated SW phase by integrating the local wavevector $k_\mathrm{SW}(x, H_\mathrm{eff}(x))$ along the propagation path (Fig.~\ref{fig:Concept}b-d), enabling rapid evaluation of the purely magnonic phase contribution for the device design. 
An analytical lumped-element transducer model~\cite{Kohl_ModellingSW,Vanderveken_Lumped,Vlaminck_2010,Bailleul_Propagating,Erdelyi_2025_Design_rules} describes the transducers as equivalent circuit elements to capture SW excitation, impedance matching, and electromagnetic crosstalk, but does not include RF feed lines or the local change in $\Heff(x)$. 
Finally, a hybrid circuit-micromagnetic simulation framework~\cite{Erdelyi_2025_Design_rules} combines transducer electromagnetic fields with full 3D micromagnetic simulations (MuMax$^3$~\cite{Vansteenkiste2011}), including the static stray field of the soft magnet, and was used for the final validation against the measured RF transmission characteristics.

\section{Operation of the Hybrid Device}
\label{sec:Operation}
\subsection{On-Chip Characterization of Hybrid Chip}
\label{ssec:ProberCharacterization}

\begin{figure}[ht]
    \centering
    \includegraphics[width=\linewidth]{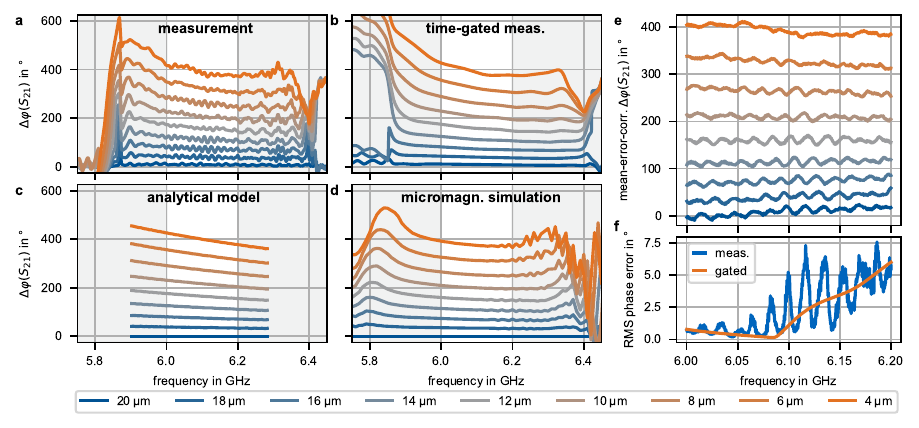}
    \caption{\textbf{Validation of measurement results, phase state diagram, and RMS phase error.} Measured a) and time-gated b) phase change $\Delta\varphi(S_{21})$ of the magnonic phase shifter for different distances $d_\mathrm{m}$ of the micromagnet to the SW chip. The reference phase is measured at zero applied DC voltage, equivalent to $d_\mathrm{m}=$\SI{20.7}{\micro\meter}. We achieve a maximum phase change $\Delta\varphi(S_{21})$ between \SI{375}{\degree} (\SI{6.2}{\giga\hertz}) and \SI{424}{\degree} (\SI{6}{\giga\hertz}) for the maximum cantilever displacement at $d_\mathrm{m}=\SI{4}{\micro \meter}$. The experiment is compared to our analytical model in c) and to micromagnetic simulations shown in d). The analytical model and micromagnetic simulations include the measured cantilever tilt. The agreement between the simulated and time-gated data is better than \SI{35}{\degree} in absolute value within a \SI{200}{\mega\hertz} span. e) The average-phase-error-corrected state diagram of the magnonic phase shifter is the basis for calculating f) the root-mean-square (RMS) phase error of less than \SI{8}{\degree} across all cantilever positions.}
    \label{fig:RF1_PhaseChange_combi}
\end{figure}

A detailed characterization of the magnonic phase shifter was performed in a wafer prober with an integrated electromagnet at a center frequency of $f_\mrm{c}=\SI{6.1}{\giga\hertz}$ and a constant global bias field of $\mu_0\Hext=\SI{152}{\milli\tesla}$. 
The hybrid chip was evaluated by 2-port measurements at an input power of \SI{-15}{\dBm}, with the micromagnet-to-chip distance $\dm$ varied from \SI{20}{\micro\meter} to \SI{4}{\micro\meter}. 
Ground-signal-ground (GSG) probes with the calibrated reference plane at the probe tips were used for all measurements. 
The signal at $\dm=\SI{20.7}{\micro\meter}$, i.e., without DC voltage supply, served as the reference (\SI{0}{\degree}) state for calculating the phase shift $\Delta\varphi(S_{21})$, shown in Fig.~\ref{fig:RF1_PhaseChange_combi}a, in which a \SI{200}{\mega\hertz} bandwidth around $f_\mrm{c}$ is highlighted. 
Within this band, the maximum phase shift reaches \SI{375}{\degree} at \SI{6.2}{\giga\hertz} and \SI{424}{\degree} at \SI{6}{\giga\hertz} for $\dm=\SI{4}{\micro\meter}$ or about \SI{20}{\volt} DC bias for the cantilever. 
The frequency dependence of $\Delta\varphi_\mrm{max}(S_{21})$ originates from the non-linearity of the SW dispersion with frequency at constant $\Hext$ and becomes more pronounced for shorter wavelengths in the backward-volume geometry~\cite{Kalinikos1986,Stancil2009}. 
This effect results in phase shifts of about \SI{590}{\degree} near \SI{5.9}{\giga\hertz} before the excitable wavevector range is defined by the CPW transducer geometry (Supplementary Information Sec.~\ref{ssec:A_MagnonicChip}).

To evaluate the intrinsic device response, a time-gating post-processing scheme (Supplementary Information Sec.~\ref{ssec:A_TimeDomainGating}) was applied, as described in~\cite{Davidkova2025,Kohl_ModellingSW}. 
After suppressing electromagnetic crosstalk and multi-path SW propagation, the time-gated phase response in Fig.~\ref{fig:RF1_PhaseChange_combi}b shows a smooth variation. 
The experimental results are compared with the semi-analytical phase-shifter model and the hybrid circuit--micromagnetic simulation framework in Fig.~\ref{fig:RF1_PhaseChange_combi}c,d. 
Experimentally determined geometrical parameters, including micromagnet alignment, micromagnet-to-YIG spacing, and cantilever tilt, were incorporated without additional fitting parameters (Supplementary Information Sec.~\ref{ssec:A_MEMScharacterization}). 
The simulations reproduce the measurements with a maximum deviation below \SI{35}{\degree} or \SI{8.2}{\percent} relative to the absolute measured $\Delta\varphi$ at \SI{6}{\giga\hertz} and $\dm=\SI{4}{\micro\meter}$.

Beyond the phase-shift performance, additional technical characteristics of the hybrid chip were evaluated and are detailed in Supplementary Information Sec.~\ref{ssec:A_RF_Sparams_Phase}--\ref{ssec:A_PowerHandling}. 
The average-phase-error-corrected phase diagram is shown in Fig.~\ref{fig:RF1_PhaseChange_combi}e and shows an improved phase-frequency flatness with slight ripples as compared to the raw measurement data.  
Based on the average-phase-error-corrected phase diagram, the device achieves an RMS phase error below \SI{8}{\degree}, reduced to \SI{6}{\degree} after time-gating (Fig.~\ref{fig:RF1_PhaseChange_combi}f). 
At $f_\mrm{c}=\SI{6.1}{\giga\hertz}$, the phase-voltage sensitivity is \SI{20.3}{\degree\per\volt} with a relative linearity deviation of \SI{8.5}{\percent} when actuating the MEMS cantilevers using a look-up table (Supplementary Information Sec.~\ref{ssec:A_MEMScharacterization}).
All data presented in this work were collected with the applied DC voltage to the cantilever ramped up step by step using a look-up table.
The look-up table uses a fitted quadratic function to represent the measured data with an $R^2$ of better than 0.998.
The transmission and reflection coefficients are approximately \SI{-22}{\decibel} and \SI{-7}{\decibel}, respectively, with variations below $\pm\SI{1.25}{\decibel}$ over the charcterized actuation range. 
The RMS transmission amplitude error remains below \SI{1}{\decibel} within a \SI{200}{\mega\hertz} span around $f_\mrm{c}$. 
Electromagnetic crosstalk between the transducers is $\SI{-55}{\decibel}\pm\SI{1}{\decibel}$ within $\pm\SI{200}{\mega\hertz}$ around $f_\mrm{c}$, corresponding to a signal-to-noise ratio of \SIrange{29.5}{34.5}{\decibel} between the transmitted SW signal and the crosstalk. 
Across the characterized actuation range, $f_\mrm{c}$ shifts by at most \SI{24.8}{\mega\hertz}, while the 3 dB bandwidth $\BWthreeDB$ varies by $\pm\SI{33}{\mega\hertz}$ around a mean value of \SI{134}{\mega\hertz}. 
The time-gated group delay varies by $\pm\SI{1.5}{\nano\second}$ around \SI{28}{\nano\second} for a \SI{10}{\mega\hertz} delay aperture. 
Power-dependent measurements yield a \SI{1}{\decibel} compression point of $\PdB=\SI{-11}{\dBm}$. 
An additional functionality test under swept bias field $\Hext$ confirms that local tunability of the SW dispersion is essential for maintaining the operating principle of the magnonic phase shifter (Supplementary Information Sec.~\ref{ssec:A_FunctionalityTest}).

\subsection{Operating Point Tunability and Multi-Channel Operation}
\label{ssec:Tunability_Comparison}

\begin{figure}[ht]
    \centering
    \includegraphics[width=\linewidth]{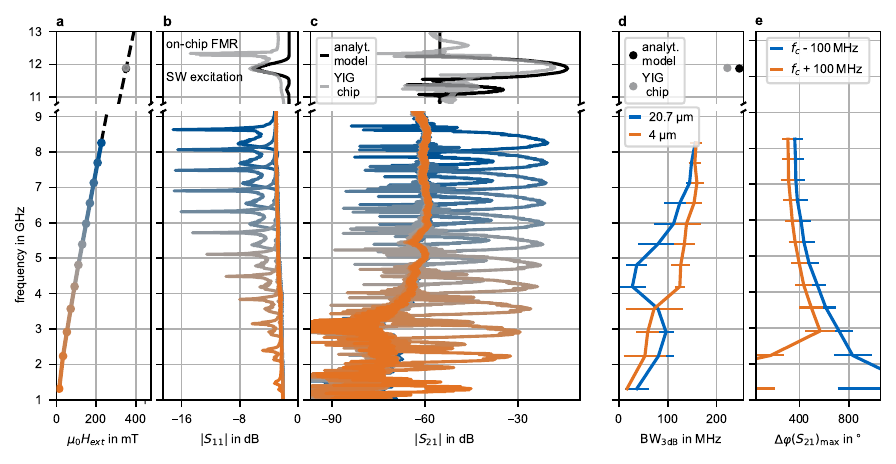}
    \caption{\textbf{Tunability of center frequency with global field and comparison of multiple channels.} a) The center frequency of the phase shifter can be tuned via $\Hext$ with a slope of about \SI{28}{\giga\hertz\per\tesla}. Magnitude of b) the forward reflection $S_{11}$ and c) the forward transmission $S_{21}$ at the highest cantilever position for varying $\mu_0\Hext$ from \SIrange{13}{226}{\milli\tesla}. With increasing frequency, the impedance matching between the SW transducer and the SW-induced radiation resistance (load impedance) improves, leading to a higher RL. As a result, the conversion efficiency from the electromagnetic system to the SW system improves, and so does the transmission. d) The mean $\BWthreeDB$ of the three characterized transmission channels on the hybrid chip is about \SI{160}{\mega\hertz} at \SI{8.2}{\giga\hertz}, while the reference measurement before the flip-chip process (gray) shows \SI{220}{\mega\hertz} at around \SI{12}{\giga\hertz} e) The max. achievable phase shift $\Delta\varphi(S_{21})_\mrm{max}$ within a \SI{200}{\mega\hertz} span around $f_\mrm{c}$ reaches about \SI{600}{\degree} at $f_\mrm{c}\approx\SI{3}{\giga\hertz}$. The black graphs in a)-d) are extracted from the lumped-element model and show excellent agreement with the reference measurement.}
    \label{fig:Hero8_VH_Sparameters_BW_PhiMax}
\end{figure}

One major advantage of spin-wave devices is that the center frequency or operational point can be tuned over a wide range by adjusting the global magnetic bias field $\Hext$. 
By varying $\mu_0\Hext$ from \SIrange{13}{226}{\milli\tesla}, the center frequency can be shifted between \SI{1}{\giga\hertz} and \SI{8.2}{\giga\hertz} with a slope of about \SI{28}{\giga\hertz\per\tesla}~\cite{Stancil2009}, as shown in Fig.~\ref{fig:Hero8_VH_Sparameters_BW_PhiMax}a.
The corresponding $S_{11}$ and $S_{21}$ are shown in Fig.~\ref{fig:Hero8_VH_Sparameters_BW_PhiMax}b-c and follow the FMR frequency with an offset of approximately \SI{300}{\mega\hertz} below $f_\mrm{FMR}$, as expected from the backward volume dispersion relation.
Starting around \SI{3}{\giga\hertz}, the excitation of propagating SWs is efficient enough to clearly elevate the transmitted SW signal out of the electromagnetic crosstalk.
The thereby reduced intermodulation between the two signal contributions allows for reliable bandwidth and phase-shift extraction.
The upper experimental limit of \SI{226}{\milli\tesla} was set by the available electromagnet. 
In an application, an integrated permanent magnet of a few millimeters in size can readily provide comparable or higher fields.
To extend the range of investigation, the upper parts of Fig.~\ref{fig:Hero8_VH_Sparameters_BW_PhiMax}a-c show the equivalent data extracted from the lumped-element model ($\mu_0\Hext=\SI{350}{\milli\tesla}$, $f_\mrm{c}=\SI{11.89}{\giga\hertz}$) and a reference measurement of the magnonic YIG chip before the flip-chip process ($\mu_0\Hext=\SI{349}{\milli\tesla}$, $f_\mrm{c}=\SI{11.87}{\giga\hertz}$).
Except for the additional detection of on-chip FMR underneath the RF feed lines on the YIG chip compared to the model used only for the SW transducers, the match between the data sets is strong, which is confirmed for a broad span of frequencies in Supplementary Information Sec.~\ref{ssec:Tunability_Comparison}.
The agreement between the predicted behavior and the reference measurement confirms that the phase shifter is also suitable for RF applications well beyond \SI{10}{\giga\hertz}. 

The hybrid chip features three independent, parallel operating channels, which were compared over the tested bias-field range to demonstrate multi-channel operability. 
The mean 3 dB bandwidth ($\BWthreeDB$) and maximum available phase shift are summarized in Fig.~\ref{fig:Hero8_VH_Sparameters_BW_PhiMax}d-e. 
The $\BWthreeDB$ increases with $f_\mrm{c}$ and $\Hext$, reaching about \SI{160}{\mega\hertz} around \SI{8.2}{\giga\hertz} for the measured data and \SI{180}{\mega\hertz} for the time-gated data (Supplementary Information Sec.~\ref{ssec:A_Variation_fc_IL_BW}).
However, the reference measurement of the magnonic YIG chip before flip-chip shown in the upper part of Fig.~\ref{fig:Hero8_VH_Sparameters_BW_PhiMax}d clearly shows that due to improving impedance matching and a steeper dispersion relation (Supplementary Information Sec.~\ref{ssec:A_MagnonicChip}), a realistic SW transmission channel allows for bandwidths around \SI{220}{\mega\hertz} at $f_\mrm{c}\approx\SI{12}{\giga\hertz}$.
This value is slightly overestimated by the channel's analytical lumped-element model, with the simplifications explained in Supplementary Information Sec.~\ref{ssec:A_Model_Description_And_Usage} and Sec.~\ref{ssec:A_Tunability_Analytical}.
In Fig.~\ref{fig:Hero8_VH_Sparameters_BW_PhiMax}e, the maximum phase shift $\Delta\varphi(S_{21})_\mrm{max}$ (lowest cantilever position) is shown at the edges of a \SI{200}{\mega\hertz} bandwidth around $f_\mrm{c}$.
The difference in $\Delta\varphi(S_{21})_\mrm{max}$ at the band edges represents about twice the peak absolute phase error and is almost constant above $\mu_0\Hext\approx\SI{100}{\milli\tesla}$ or $f_\mrm{c}\approx\SI{4}{\giga\hertz}$.
In Fig.~\ref{fig:Hero8_VH_Sparameters_BW_PhiMax}d-e, the standard deviation between the three investigated channels is provided as errorbars and confirms the reproducibility of the used fabrication and assembly technology.

\subsection{System-Level Operation}
\label{sec:StandaloneOperation}

\begin{figure}[t!]
    \centering
    \includegraphics[width=0.9\linewidth]{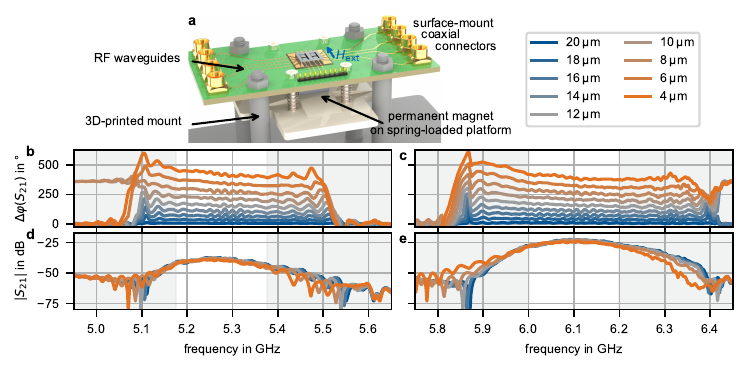}
    \caption{\textbf{System-level evaluation of the magnonic phase shifter}. a) The hybrid chip is embedded and wire-bonded to a PCB. The external magnetic bias field is provided by a permanent magnet mounted on a spring-loaded platform beneath the PCB, and the platform is characterized using coaxial cables. The phase shift $\Delta\varphi(S_{21})$ b) and magnitude d) of $S_{21}$ are compared to the equivalent characterization in c) and d) of the bare hybrid chip when evaluated in the wafer prober. The reduced overall impedance matching in the non-optimized system-level configuration reduces the transmitted power level by about \SI{15}{\decibel} and, thus, the achievable bandwidth. Nevertheless, the phase-shift characteristics are not impaired and yield the same results for both evaluated configurations. In plots b)-e), the bright area highlights a \SI{200}{\mega\hertz} bandwidth around $f_\mrm{c}$.}
    \label{fig:StandaloneComparison}
\end{figure}

For demonstrating the system-level operability of the magnonic RF phase shifter, we used a carrier PCB and wire-bonded the CPW feed lines on the magnonic chip to the CPW traces on the board, as shown in Fig.~\ref{fig:Concept}h.
The carrier PCB is a 4-layer FR4 board with four \SI{50}{\ohm} CPW input and output traces to/from surface-mount SMP coaxial connectors.
A rendered and simplified visualization is provided in Fig.~\ref{fig:StandaloneComparison}a.
A cube-shaped permanent magnet with a side length of \SI{10}{\milli\meter} is mounted on a spring-loaded platform beneath the carrier PCB to adjust $\Hext$.
The bias field can be tuned to $f_\mrm{c}=\SI{6.1}{\giga\hertz}$ only for two of the three channels simultaneously, which is a limitation of the simplistic method of using a permanent magnet for biasing. 
Further optimization is needed to enable a realistic multi-channel device behavior.
A summary of the RF characteristic of the carrier PCB is provided in the Supplementary Information Sec.~\ref{ssec:A_CarrierBoard}.

The measurements discussed in Sec.\ref{sec:Operation} are repeated and compared with side-by-side regarding the phase shift $\Delta\varphi(S_{21})$ in Fig.~\ref{fig:StandaloneComparison}b-c and the forward transmission $S_{21}$ in Fig.~\ref{fig:StandaloneComparison}d-e.
The center frequencies of \SI{5.25}{\giga\hertz} (permanent magnet) and \SI{6.1}{\giga\hertz} (electromagnet in prober) are reasonably close to each other, so that a similar behavior can be expected.
Regarding the achievable phase shift, there is no difference between platform-level analysis and characterization in the RF wafer prober, and in both cases, we achieve $\Delta\varphi_\mrm{max}\ge\SI{400}{\degree}$.
However, the available bandwidth is reduced by about \SI{80}{\mega\hertz}, due to an impedance mismatch between the hybrid chip and the non-optimized carrier PCB.
This system-level characterization provides clear evidence that, due to the very small active area of the magnonic phase shifter, the field from the small permanent magnet is sufficiently homogeneous to replace bulky, power-hungry electromagnets.

\section{Discussion}
\label{sec:Discussion}

While local modification of the SW dispersion via static magnetic field inhomogeneities~\cite{Cocconcelli_Tuning_2024} or current-driven Oersted fields~\cite{Kostylev2007,Hansen_Demidov_Demokritov_2009} has been demonstrated previously, to our knowledge, this work presents the first realization of a magnonic device in which this modification is dynamically and reversibly tuned in real time via MEMS-actuated micromagnet displacement.
This functionality extends tunable magnonic RF devices beyond filters based on globally tuned bias fields~\cite{Devitt_2026_SW_filer_6G,Du_Olsson_2024_Filters_zero_power} toward more advanced concepts such as analog phase shifters or time-delay units. 
In contrast to SW filters, which primarily require shifting the global transmission band, the presented device relies on a spatially confined and symmetric modification of the local effective field $\Heff(x)$ while maintaining stable transmission characteristics. 
This additional level of control substantially increases the device concept's complexity and highlights the capability of local dispersion engineering for RF signal processing using propagating SWs.

A key technological aspect of the presented approach is that phase tuning is achieved solely through magnetic stray fields. 
The global operational point or center frequency is defined by the external bias field $\Hext$, whereas the phase shift is controlled by the local stray field of micron-sized soft magnets integrated on piezoelectric MEMS cantilevers. 
As a result, no current-driven electromagnets or Oersted-field conductors are required for device operation, as is in principle demonstrated by the system-level integration of the phase shifter in Sec.~\ref{sec:StandaloneOperation}. 
Compared to concepts relying on current-induced magnetic fields~\cite{Hansen_Demidov_Demokritov_2009}, this architecture strongly reduces static power dissipation and avoids large ohmic losses.
In the presented implementation, the piezo's leakage current is in the order of \SI{1}{\nano\ampere} at the maximum applied voltage of \SI{20}{\volt}, resulting in a static loss of about \SI{20}{\nano\watt}.
For dynamic operation close to the mechanical resonance, the dissipated power is calculated via $P=(V/Q)^22\pi f C_0\tan\delta\approx\SI{12}{\nano\watt}$ using the measured resonance frequency $f=\SI{11.9}{\kilo\hertz}$, quality factor $Q=\num{193}$, static capacitance $C_0=\SI{1.4}{\nano\farad}$, and a typical dielectric loss tangent $\tan\delta\approx\num{0.01}$ for the used piezoelectric material.
In practice, power consumption is therefore dominated by the actuator driver electronics, which have static power dissipation ranging from tens of microwatts~\cite{LTC1050} to the mW range~\cite{FAN8831,OPA2188}.
Even including dynamic charging and discharging losses of the piezoelectric capacitance, the overall power consumption is expected to remain in the sub-mW regime, and, together with the absence of continuously driven electromagnets, highlights the potential of the presented concept for scalable and tunable ultra-low-power RF hardware. 

\begin{table*}[t!]
\centering
\scriptsize
\setlength{\tabcolsep}{3pt}
\renewcommand{\arraystretch}{1.35}
\caption{Comparison of prototypical and commercial tunable RF phase shifters based on different technologies. Besides conventional RF performance metrics, a footprint-normalized figure of merit (FOM) is introduced to assess the maximum phase shift per unit insertion loss and per unit channel area. The presented tunable magnonic phase shifter combines a compact footprint and scalable low-power operation, while offering significant optimization potential through device scaling and improved SW propagation geometries.}
\label{tab:FOM_Quantitites}
\resizebox{\textwidth}{!}{%
\begin{tabular}{m{1.9cm}m{2cm}m{2.5cm}m{2.5cm}m{2.5cm}m{2.5cm}m{1.4cm}m{1.4cm}m{2.2cm}m{2.5cm}}
\textbf{Reference} & \textbf{frequency range\newline in GHz} & \textbf{insertion\newline loss\newline in dB} & \textbf{phase-voltage-sensitivity\newline in $^\circ$/V} & \textbf{phase-voltage-linearity dev.\newline in \% $^a$} & \textbf{max.\newline phase\newline shift in $^\circ$} & \textbf{channel\newline footprint\newline in mm$^2$} & \textbf{device\newline footprint\newline in mm$^2$} & \textbf{FOM$^b$\newline in\newline $^\circ$/(dB$\times$mm$^2$)}\\
\toprule
This work & 3--8 shown,\newline $>10$ possible & 22 ($>$4GHz) & 20.3 & $8.5\%$ & 400 (4--8GHz) &  0.5 shown,\newline 0.15 possible$^c$ & 100 & 36$^d$, 121$^c$, 267$^e$\\
\hline
YIG/BST\newline\cite{Ustinov_Kolkov_Nikitin_Kalinikos_Fetisov_Srinivasan_2011} & 3.5--8 shown & 15 @ 6GHz & 0.36~@~4.9GHz\newline 0.63~@~7.65GHz & $\sim13\%$ @ 5GHz\newline $<10\%$ @ 7.7GHz & $360^\circ$ @ 5GHz\newline $600^\circ$~@~7.7GHz & 14.2 & $\sim1000$ & 1.7 @ 5GHz\newline 2.8 @ 7.7GHz \\
\hline
YIG \newline + pulsed\newline current\cite{Hansen_Demidov_Demokritov_2009} & 4 shown & 23--27 @ 4GHz & 306$^\circ$/A$^{f}$ & $\sim12\%$ $^f$ & $360^\circ$ @ 4GHz $^g$ & 7.2 & $>22.5$ & 2.1 \\
\hline
MEMS\newline + DTML\cite{Kumar_DTML_MEMS_2022} & 0--18 shown & typ. $<2$ & 3 & $\sim10\%$ & 60 & $>4$ & $\sim5.6$ & 7.5\newline (0-18GHz)\\
\hline
BST\newline in CPW\cite{Acikel_PhaseShifter_2002} & 0--10 shown & typ. 2.2~$<$10GHz\newline max. 3 @ 10GHz & 13 & $<10\%$ & 240 & $\sim20$ & 61 & 1.7\newline (0-10GHz) \\
\hline
Liquid crystals\cite{Li_LCphaseShifter_2019} & 54--65 & $<4$ & 18 & $<10\%$ & 180 & $\sim4.2$ & 240 & 0.2\newline (54-65GHz) \\
\toprule
CMD297\newline\cite{Qorvo_CMD297} & 5--18 & typ. 3,\newline max.~12~@~5GHz & 24 & $<5\%$ & 850 @ 5GHz\newline 555 @ 7GHz\newline 250 @ 12GHz & 0.16$^h$ & 3.17 & 408 @ 5GHz\newline 500 @ 7GHz\newline 520 @ 12GHz \\
\hline
HMC247\newline\cite{analogdevices_HMC247} & 5--18 & typ. 8\newline max. 12 $<$10GHz & typ. 40,\newline (max. 45) & $\sim$11\% @ 6GHz\newline $\sim$15\% @ 18GHz & 400 ($<$10GHz)\newline 200 ($>$10GHz) & 0.18$^h$ & 3.68 & 308 @ 5GHz\newline 350 @ 10GHz\newline 222 @ 18GHz \\
\hline
SCPHS-180+~\cite{MiniCircuits_SMD_SCPHS-180} & 0.09--0.18 & typ. 2.2\newline max. 5 & 50 & $\sim15\%$ & 500 & 22$^h$ & 450 & 10.4$^h$ \\
\hline
\multicolumn{9}{l}{$^a$ normalized to the absolute full-span phase shift; $^b$ FOM = max. phase shift per channel area per insertion loss; $^c$ with dedicated MEMS layout;}\\
\multicolumn{9}{l}{$^d$ as demonstrated; $^e$ with dedicated MEMS layout and in DE-geometry according to~\cite{Kohl_identification}; $^f$ exception: phase-current-sensitvity;}\\
\multicolumn{9}{l}{$^g$ @ \SI{1}{\ampere} pulsed current; $^h$ only housing footprint available for~\cite{Qorvo_CMD297,analogdevices_HMC247,MiniCircuits_SMD_SCPHS-180}. We assume a channel area of \SI{5}{\percent} of the device footprint for the FOM.}\\
\multicolumn{9}{l}{YIG: Yttrium-Iron-Garnet, BST: Barium-Strontium-Titanate, DTML: distributed transmission line}
\end{tabular}%
}
\end{table*}

The demonstrated phase shifter further provides a unique combination of frequency agility and stable phase-shift functionality. 
The center frequency can be continuously tuned via the global bias field $\Hext$, while the local phase steering mechanism remains unchanged. 
Consequently, the device maintains an almost constant maximum phase shift $\Delta\varphi_\mrm{max}>\SI{400}{\degree}$ above approximately \SI{4}{\giga\hertz}, enabling operation across different frequency bands using the same device architecture. 
This property differs fundamentally from many electronically tuned RF phase shifters, where the maximum phase shift strongly depends on frequency. 
We experimentally demonstrate operation from \SIrange{3}{8}{\giga\hertz}, while the agreement between measurements, models, and simulations suggests scalability to frequencies beyond \SI{10}{\giga\hertz}. 
This capability opens the prospect of frequency-agile RF systems in which the same phase-shifter architecture can be deployed across multiple frequency bands by adjusting only the static bias field.

The demonstrated phase shifter further benefits from an extremely small core phase-shifting area below \SI{0.02}{\milli\meter\squared}, enabled by the short SW wavelengths at GHz frequencies. 
Such compact active regions reduce the required field-biased volume and relax the demands on field homogeneity, facilitating the use of small in-package permanent magnets or sub-millimeter integrated permanent magnets~\cite{Gojdka2022,Cocconcelli_Tuning_2024}.
Using embedded permanent magnets recently enabled the demonstration of standalone integrated magonic devices~\cite{Cocconcelli2025} and shows a pathway to reduce the total device footprint and volume.
At the same time, the platform inherently supports multiple independent channels within a single package. 
Although the present demonstrator prioritizes device validation, substantial footprint reduction is feasible through dedicated co-design of the magnonic and MEMS subsystems. 
Using a refined MEMS layout, the channel footprint is expected to decrease from the demonstrated \SI{0.5}{\milli\meter\squared} to approximately \SI{0.15}{\milli\meter\squared}, while future advances in transducer scaling, alignment accuracy, and fabrication technology offer a realistic path toward even denser integration.

The current device performance and projected optimization potential are summarized in Tab.~\ref{tab:FOM_Quantitites}. 
While the proof-of-concept implementation already achieves phase-voltage sensitivities and linearity comparable to commercial analog phase shifters, the insertion loss can be further reduced by improving microwave-to-SW coupling efficiency, optimizing impedance matching~\cite{Erdelyi_2025_Design_rules,Kohl_identification}, and increasing the YIG thickness~\cite{Davidkova_FSL_2025}. 
An additional major opportunity arises from switching to unidirectional propagating SWs in the Damon--Eshbach configuration~\cite{Kohl_identification}, where integrated microscale SW transducers have already demonstrated insertion losses below \SI{10}{\decibel} even in sub-micrometer-thick YIG films. 
Implementing this geometry in the presented concept would require micron-scale hard magnets on the MEMS cantilevers, and offers a realistic route toward substantially improved RF performance while maintaining the advantages of local static-field tuning. 
In parallel, dedicated co-design of the hybrid chip and the surrounding RF environment will become increasingly important for highly integrated matched systems with broad bandwidths.

The presented device demonstrates that local dispersion engineering of propagating SWs can provide a technologically relevant route toward compact, frequency-agile, and ultra-low-power RF phase shifters. 
The combination of MEMS actuation, static-field control, and propagating spin-wave signal processing establishes a scalable platform that extends the functionality of integrated magnonic RF devices beyond tunable filters. 
Given its small active area, multi-channel capability, frequency flexibility, and compatibility with established RF packaging technologies, we envision future integrated magnonic phase-shifter banks and reconfigurable RF front-ends for applications ranging from phased-array beam steering to high-throughput wireless communication systems.

\section{System Fabrication and Methods}
\label{sec:Fabrication_Methods}
\subsubsection*{Magnonic Chip}
The magnonic chip is a \SI{800}{\nano\meter} thick Yttrium-Iron-Garnet (YIG) film grown via liquid-phase epitaxy (LPE) on a \SI{500}{\micro\meter} thick Gadolinium-Gallium-Garnet (GGG) substrate, cut to a chip size of \SI{10}{\milli\meter}~$\times$~\SI{10}{\milli\meter}.
As measured by ferromagnetic resonance spectroscopy \cite{Kalarickal2006_FMR}, the YIG film has an effective magnetization of $M_\mrm{eff}=\SI{159.8}{\kilo\ampere\per\metre}$ and a gyromagnetic ratio of $\gamma=2\pi \cdot \SI{28.06}{\giga\hertz\per\tesla}$. 
In addition, a standard Gilbert damping value of $\alpha = 2\cdot 10^{-4}$ is assumed \cite{Dubs2020_Damping}.
The transducer metallization of the magnonic chip consists of $\mathrm{SiO_2}/\mathrm{Cr}/\mathrm{Cu}/\mathrm{Cr}/\mathrm{Au}$ with thicknesses $\SI{10}{\nano\meter}/\SI{10}{\nano\meter}/\SI{1000}{\nano\meter}/\SI{10}{\nano\meter}/\SI{100}{\nano\meter}$.
The utilized metal stack minimizes YIG degradation due to the $\mathrm{SiO_2}$ barrier and facilitates bond adhesion.
The SW transducers are fabricated in a coplanar-waveguide (CPW) configuration (also ground-signal-ground, GSG) with a \SI{2}{\micro\metre} width of the lines, a gap spacing of \SI{3}{\micro\metre}, and a length of \SI{160}{\micro\metre}. 
The transducer-to-transducer spacing (center-to-center) is \SI{98}{\micro\metre}.
A detailed description of the transducer layout is given in the Supplementary Information Sec.~\ref{ssec:A_MagnonicChip}.

\subsubsection*{MEMS Chip}
The MEMS device, originally presented in \cite{ghisi_piezo-mems_2021}, consists of an array of unimorph piezoelectric cantilevers. 
Each cantilever comprises two \SI{2}{\micro\meter}-thick Lead Zirconate Titanate (PZT) patches deposited on top of a \SI{10}{\micro\meter}-thick silicon layer and is connected to the adjacent cantilever by a silicon joint. 
The device has been functionalized by depositing micromagnets on the cantilever tips. 
Each magnet has a lateral dimension of $\SI{170}{\micro\meter}~\times~\SI{98}{\micro\meter}$, whereas the latter is equivalent to the center-to-center spacing between the input and output transducers of the SW channels on the magnonic chip to achieve maximum phase shift with minimal change of the transmission band.
The micromagnets show a saturation magnetization $M_\mrm{S}\approx\SI{640}{\kilo\ampere\per\meter}$.
The micromagnets are fabricated in three steps: (i) a bilayer of a lift-off resist and AZ5214E is structured via optical lithography; (ii) a $\mathrm{Cr}/\mathrm{Ni}_{75}\mathrm{Fe}_{20}\mathrm{Mo}_{5}/\mathrm{Cr}$ multilayer is deposited by sputtering to a total thickness of \SI{695}{\nano\meter}, of which \SI{640}{\nano\meter} is magnetic material; and (iii) the magnets are defined by lift-off of the photoresist mask. 
The silicon joint between each pair of cantilevers is removed using focused-ion-beam (FIB) milling to release the two structures and enable large displacements.
Supplementary Information Sec.~\ref{ssec:A_MEMScharacterization}) provides an overview of the whole process.

\subsubsection*{Hybrid System assembly}
The magnonic chip and the MEMS chip are connected via flip-chip bonding. 
First, a set of gold stud bumps is formed on the magnonic chip using a wire bonder. 
Each interconnect consists of a gold stud bump formed by ball bonding and has a height of approximately \SI{50}{\micro\meter}. 
The two chips are then aligned with a nominal accuracy of \SI{5}{\micro\meter} and bonded via thermocompression at \SI{360}{\celsius} for \SI{150}{\second}.
After flip-chip bonding, the piezoelectric actuators are conditioned by cyclic excitation to stabilize the displacement response and minimize hysteresis. 
We applied \num{750} voltage ramps from \SI{0}{\volt} to \SI{20}{\volt} and back, in steps of \SI{0.1}{\volt}.

\subsubsection*{RF Characterization}
The main RF characterization of the hybrid chip was performed in a wafer prober with an integrated electromagnet using a vector network analyzer (VNA).
While the magnonic/MEMS chip is glued and bonded (DC supply for the cantilevers) to the carrier PCB mounted inside the electromagnet, RF ground-signal-ground (GSG) probes with a pitch of \SI{150}{\micro\meter} were used to contact the coplanar waveguides (CPWs) landing (later bonding) pads on the magnonic chip during the main characterization.
The VNA's measurement plane was calibrated to the tips of the GSG probes using an industry-standard Through-Open-Short-Match (TOSM) calibration routine.
RF measurements around $f_\mrm{c}=\SI{6.1}{\giga\hertz}$ were conducted with a span of \SI{4.2}{\giga\hertz}, a frequency step size $\Delta f=\SI{125}{\kilo\hertz}$, an intermediate-frequency bandwidth (IFBW) or resolution bandwidth of \SI{1}{\kilo\hertz}, and without averaging or smoothing.
Additional measurements for varying center frequencies and the pure electrical characterization were conducted from \SIrange{1}{18}{\giga\hertz}, $\Delta f=\SI{1}{\mega\hertz}$, and otherwise unchanged settings.
For the standalone characterization, the hybrid chip was wire-bonded to the carrier PCB, and the whole platform was characterized using TOSM-calibrated, de-embedded coaxial cables with measurement parameters similar to those used in the main characterization.

\backmatter

\bmhead{Acknowledgments}
Financial support from the EU Horizon Europe research and innovation program, within the project “MandMEMS” (Grant No. 101070536), is gratefully acknowledged.\\
We thank STMicroelectronics for providing MEMS devices in the framework of the JRP STEAM. 
This work has been partially performed at Polifab, the micro and nanofabrication facility of Politecnico di Milano.\\

\bmhead{Author contributions}
\textbf{J.G.} designed the layout of the magnonic chip \& the RF carrier PCB, characterized the MEMS actuation as well as the hybrid magnonic/MEMS system both in the wafer prober and in the system-level configuration, and assembled the main draft of the manuscript.
\textbf{A.A.} functionalized the MEMS chip by fabricating the micromagnets on the cantilevers, characterized the magnets and the electric behavior of the MEMS cantilevers, and performed the flip-chip process.
\textbf{F.K.} refined the analytical model of the SW transducer behavior, including electromagnetic crosstalk and multi-path propagation, designed the layout of the SW transducers, and performed the characterization of the magnonic chip before flip-chip bonding.
\textbf{A.P.} performed the electromagnetic (FEMM) and micromagnetic (Mumax$^3$) simulations for the hybrid circuit-micromagnetic modeling.
\textbf{M.W.} developed the analytical model for describing the SW phase shifter functionality.
\textbf{B.H.} prepared the bare magnonic chips, fabricated the chip metallization, and was responsible for the FIB release cut.
\textbf{R.B.}, \textbf{F.M}, and \textbf{P.P.} developed the concept of the tunable magnonic phase shifter using stray fields from MEMS-displaced micromagnets.\newline   
All authors contributed to the discussion of the results and revision of the manuscript.

\bmhead{Competing interests}
The authors declare no competing interests.

\bmhead{Author information}
Johannes Greil \url{https://orcid.org/0009-0009-5693-7736}\newline
Antonio Angotti \url{https://orcid.org/0009-0006-4440-3401}\newline 
Felix Kohl \url{https://orcid.org/0009-0003-6131-1275}\newline
\'{A}d\'{a}m Papp \url{https://orcid.org/0000-0002-4286-6828}\newline
Matthias Wagner \url{https://orcid.org/0009-0000-4060-1378}\newline
Maria Cocconcelli \url{https://orcid.org/0000-0002-1201-0295}\newline
Andrea Del Giacco \url{https://orcid.org/0009-0002-4260-127X}\newline
Dieter Ferling \url{https://orcid.org/0000-0002-7581-6309} \newline
Björn Heinz \url{https://orcid.org/0000-0002-0162-1007}\newline
Federico Maspero \url{https://orcid.org/0000-0001-8220-5509}\newline
György Csaba \url{https://orcid.org/0000-0001-7592-0256}\newline
Riccardo Bertacco \url{https://orcid.org/0000-0002-8109-9166}\newline
Markus Becherer \url{https://orcid.org/0000-0002-1291-8877}\newline
Philipp Pirro \url{https://orcid.org/0000-0002-0163-8634}\newline

\newpage

\begin{appendices}

\section{Description and Application of Semi-Analytical and Hybrid Circuit-Micromagnetic Simulation Approaches}
\label{sec:A_ModelExplanation}

\subsection{Model Description and Usage}
\label{ssec:A_Model_Description_And_Usage}

\begin{figure}[ht]
    \centering
    \includegraphics[trim={0 0 0 0},clip,width=0.9\linewidth]{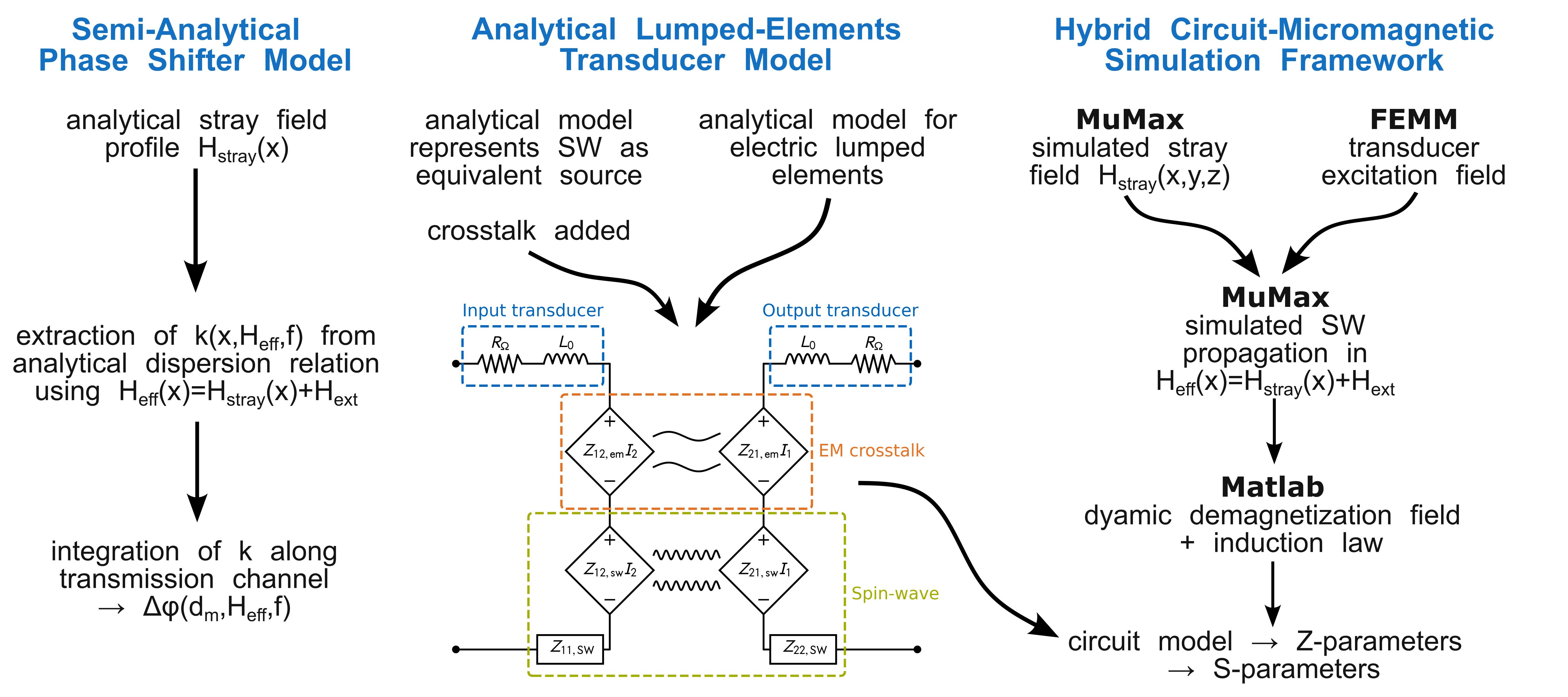}
    \caption{Schematic overview of the analytical models and the hybrid circuit--micromagnetic simulation framework. a) The semi-analytical phase-shifter model enables rapid design iterations and captures the purely magnonic contribution to the phase shift. b) The analytical lumped-element transducer model combines the electrical/electromagnetic and magnonic contributions to the coupling between the input and output transducers, and is used to design the transducer layout and evaluate impedance matching~\cite{Kohl_ModellingSW}. c) The hybrid circuit--micromagnetic simulation framework of~\cite{Erdelyi_2025_Design_rules}, extended to include the static stray field of the micromagnets, is used to validate the RF characterization.}
    \label{fig:ModelExplanation}
\end{figure}

We employed two analytical models and a hybrid circuit-micromagnetic simulation framework to design and validate the magnonic RF phase shifter, as schematically illustrated in Fig.~\ref{fig:ModelExplanation}. 
For rapid design iterations, a semi-analytical phase-shifter model was developed. 
To calculate the accumulated spin-wave phase $\varphi$, an adiabatic variation of the spin-wave wave vector $k$ without reflections is assumed during propagation through the stray field of the soft magnet. The procedure consists of the following steps:
\begin{enumerate}[label=(\arabic*), itemsep=8pt]
    \item Calculate the local spin-wave dispersion relation $f_k(x_n)$ (Sec.~\ref{ssec:Dispersion_Calculation}) at equidistant positions $x_n=n\,\Delta x$ ($n=0,\dots,N$) between the center lines of the input ($x_0=0$) and output antenna ($x_N=D$; $D=\SI{98}{\micro\meter}$ in the experiment).
    
    \item For a fixed frequency $f$, extract the corresponding spin-wave wave vector $k_n$ satisfying $f_{k_n}(x_n)=f$ from the local dispersion relation.
    
    \item Compute the accumulated spin-wave phase $\varphi(f)=\int_0^D k_x\,dx$ by numerically integrating the wave vector distribution $[k_0,k_1,\dots,k_N]$ along the propagation direction.
    
    \item Determine the phase shift $\Delta\varphi(f)$ by subtracting the reference phase $\varphi_\text{ref}$, obtained for $\dm=\SI{20}{\micro\meter}$, from the phases calculated at the remaining values of $\dm$.
\end{enumerate}

To evaluate $\Delta\varphi$ across the excitation band of the SW transducer, steps \hbox{(2)--(4)} are repeated for a frequency grid spanning the predicted operational frequency band of the device. 
The stray field of the soft magnet is included by assuming a single-domain magnetization aligned with the external bias field $\mu_0\Hext$. 
For a rectangular magnet cross-section, this approximation yields an analytical expression for the stray field $\mu_0H_\text{stray}$, as described in~\cite{Engel_Herbert2005}. 
The parameters used in the semi-analytical model are summarized in Tab.~\ref{tab:num_params_model}.

An analytical lumped-element transducer model is used to describe the RF behavior of the transducers (excluding feed lines) in terms of an impedance parameter matrix ($Z$-matrix), which can be converted into a scattering parameter matrix ($S$-matrix). 
The model is based on calculating the voltages and currents induced in the SW transducers by the oscillating stray field of propagating SWs~\cite{Vanderveken_Lumped,Vlaminck_2010,Bailleul_Propagating}. 
It was further refined in~\cite{Kohl_ModellingSW,Erdelyi_2025_Design_rules} by incorporating an approximation of the electrical self-impedance and equivalent sources representing electromagnetic crosstalk and propagating SW contributions between the transducers. 
The agreement between model and experiment reported in~\cite{Kohl_ModellingSW,Erdelyi_2025_Design_rules} supports the treatment of the SW channel as an electrically small (lumped) element, whereas conventional feed lines must be modeled separately. 
The model was used to design the SW transducer geometry and to estimate the microwave-to-SW coupling efficiency and impedance matching. 
Combined with the analytical SW dispersion relation~\cite{Kalinikos1986}, it also enables the calculation of bias-field- and frequency-dependent circuit characteristics, as demonstrated in Supplementary Information Sec.~\ref{ssec:A_Tunability_Analytical}.

\begin{table}[ht] 
\caption{Material and numerical parameters used for the semi-analytical phase shifter model.}
    \centering
    \captionsetup{width=\textwidth}
    \normalsize
    \begin{tabular}{l l l} 
        \textbf{Parameter} & \textbf{Unit} & \textbf{Value} \\
        \toprule
        Num. cos-modes: $j_\text{max}$ & - & 10 \\
        Spatial resolution: $\Delta x$ & $\si{\micro \meter}$ & 0.5 \\
        Sat. magn. soft magnet & $\si{\kilo \ampere \per \meter}$ & 640 \\
        Length soft magnet: $L_x$ & $\si{\micro \meter}$ & 98 \\
        Width soft magnet: $L_y$ & $\si{\micro \meter}$ & 170 \\
        Thickn. soft magnet: $L_z$  & $\si{\nano \meter}$ & 640 \\
    \end{tabular}
    \label{tab:num_params_model}
\end{table}

For the design and validation of the device concept, we employed the hybrid circuit-micromagnetic simulation framework introduced in~\cite{Erdelyi_2025_Design_rules}. 
For the magnonic phase shifter, the framework was extended to include the stray field $H_\mrm{stray}$ of the actuated micromagnets. 
The CPW transducers were modeled in FEMM~\cite{FEMM} assuming a reference current in the conductors. 
Using the 2D time-harmonic magnetic solver, FEMM calculates the current distribution and resulting magnetic field while accounting for eddy currents in the finite conductor cross-section. 
The calculated fields are used as excitation in full 3D micromagnetic simulations performed with mumax\textsuperscript{3}~\cite{Vansteenkiste2011}, which include the complete transducer length but exclude the substantially larger feed lines and contact pads. 
Broadband excitation was implemented using a $\mrm{sinc}$ current pulse with equal spectral amplitude in the range from \SIrange{5}{7}{\giga\hertz}.
The MEMS magnet was modeled as a static single-domain rectangular magnet and assumed to be non-resonant within the operating frequency range. 
Its magnetization and stray field were calculated in a separate mumax\textsuperscript{3} simulation, and the resulting $H_\mrm{stray}$ was imported into the dynamic simulations as a static external field. 
Transmission characteristics for different magnet distances and tilts were obtained from separate simulations using the corresponding stray-field distributions.
For comparison with the fabricated device, the simulation geometry was set up according to the measured structure, including layer thicknesses, chip-to-chip alignment, and the cantilever tilt along the SW propagation direction. 
The time-dependent magnetization was post-processed to calculate the induced voltages and equivalent two-port impedance parameters using the procedure described in~\cite{Erdelyi_2025_Design_rules}. 
The resulting impedance parameters were converted into scattering parameters and compared with the experimental data, as presented in Supplementary Information Sec.~\ref{ssec:A_ComparisonMuMax}.

\subsection{Numerical Dispersion Calculation}
\label{ssec:Dispersion_Calculation}

In the demonstrated phase shifter, the spin-wave dynamics are affected by two magnetic fields of different amplitudes: The strong global bias field $\mu_0 H_\text{ext}$ and the comparably weak stray field $\mu_0 H_\text{stray}$ of the soft magnet. 
The former determines the working point of the dispersion relation, whereas the stray field acts as a weak modulation, keeping the plane-wave character of the spin waves approximately unaffected. 
This enables the computation of the local dispersion relation $f_k(x)$ along the propagation direction $x$ in the same way as for a homogeneously magnetized film, by including the local magnetic field $\mu_0 H_\text{eff}(x)= \mu_0 \left(H_\text{ext} + H_\text{stray}(x) \right)$ as the external bias of the system.

The dispersion relation $\omega_{k,n} = 2\pi f_{k,n}$ is therefore computed following the approach by Kalinikos and Slavin \cite{Kalinikos1986}, i.e., solving the linearized Landau-Lifshitz-Gilbert equation in the plane-wave basis for spin-wave modes $n$ with different spatial profiles $\textbf{m}_{k,n}(z)$ across the film thickness $L$
\begin{equation} \label{eq:LLG}
   \text{i} \omega_{k,n} \textbf{m}_{k,n} = \textbf{m}_0 \times \hat{\bm{H}}_k \cdot \textbf{m}_{k,n}.
\end{equation}
In the experiment, the total effective field acts on the sample under some angles $\theta_\text{eff}$ and $\phi_\text{eff}$ with respect to the film normal and the propagation direction $\hat{\bm{k}}$ of the spin waves. From this field configuration, the orientation of the equilibrium magnetization
\begin{equation}
   \textbf{m}_0 = \begin{pmatrix} \cos(\phi_\text{eff}) \, \sin(\theta_0) \\ \sin(\phi_\text{eff}) \, \sin(\theta_0) \\ \cos(\theta_0)
    \end{pmatrix}
\end{equation}
and the value of the internal field $B_0=\mu_0 H_0$ can be obtained, solving the magneto-static problem introduced in~\cite{Kalinikos1986}
\begin{eqnarray}
	\begin{cases}
		H_0 \cos (\theta_0) = H_{\text{eff}} \cos (\theta_\text{eff}) - M_{\text{s}} \cos (\theta_0) \\
		M_\text{s} \cos (2 \theta_0) = 2 H_{\text{eff}} \sin \big(\theta_0 - \theta_\text{eff} \big)\,.
	\end{cases}
\end{eqnarray}
In ~\eqref{eq:LLG}, the tensor $\hat{\bm{H}}_k$ comprises all interactions in the magnetic system that are responsible for the coupling of the magnetic moments:
\begin{equation}
    \begin{split}
     &\left[ \hat{\bm{H}}_k \cdot \textbf{m}_{k,n} \right](z) \\
     &= \omega_\text{B} \hat{\bm{I}} \cdot \textbf{m}_{k,n} (z) + \omega_\text{M} \left[\hat{\bm{N}}_k \cdot \textbf{m}_{k,n}(z) - \int_0^L \left( \hat{\bm{G}}_k(z-z')\cdot \textbf{m}_{k,n}(z') \right) \, \text{d}z'  \right],
    \end{split} 
\end{equation}
where $\omega_\text{B} = \gamma B_0$ and $\omega_\text{M} = \gamma \mu_0 M_\text{S}$. In the case of an magnetically isotropic medium as YIG, this includes the contribution of the exchange
\begin{equation}
     \hat{\bm{N}}_k = \lambda_\text{ex}^2 \left( k^2 - \frac{d^2}{dz^2}  \right) \hat{\bm{I}} \, , \label{eq:exchange_operator}
\end{equation}
as well as the dipolar
\begin{equation}
    \begin{split}
        \hat{\bm{G}}_k(z-z') & = (\hat{\textbf{z}} \otimes \hat{\textbf{z}}) \, \delta(z-z') \\
        & + (\hat{\textbf{k}} \otimes \hat{\textbf{k}} - \hat{\textbf{z}} \otimes \hat{\textbf{z}}) \, P(z-z') \\
        & + i (\hat{\textbf{k}} \otimes \hat{\textbf{z}} + \hat{\textbf{z}} \otimes \hat{\textbf{k}}) \, Q(z-z') \label{eq:dipole_operator}
    \end{split}
\end{equation}
interaction. Here, $\lambda_\text{ex}$ denotes the exchange length of the material whereas $P(z) = \tfrac{k}{2} \, e^{-k |z|}$ and $Q(z) = \text{sgn}(z) P(z)$ are the two matrix elements of the dipolar Green's function \cite{Guslienko2011}.

The integro-differential equation~\eqref{eq:LLG} is supplemented by boundary conditions, defining the alignment of the spins at both film interfaces $z=0,L$ . Following~\cite{Kalinikos1986}, we assume that the surface spins are unbounded 
\begin{equation}
    \frac{d\textbf{m}}{dz} \bigg \vert_{z=0,L} = 0 \, ,
\end{equation}
as this allows to decompose the mode profiles into a series of normalized cos-functions 
\begin{equation}
    \textbf{m}_{k,n} = \sum_j \textbf{m}_{k,n}^{(j)} \cdot \alpha_j \, \cos(j \pi z/L) \quad \text{with} \quad \alpha_j = \begin{cases} 1, \, j=0 \\
    \sqrt{2}, \, j>0
    \end{cases} \, .  
\end{equation}
When inserting this cos-representation into~\eqref{eq:LLG}, the eigenvalue problem is mapped to an infinite-matrix-vector system for the expansion coefficients $\textbf{m}_{k,n}^{(j)}$ 
\begin{equation} \label{eq:LLG_cos_basis}
   \text{i} \omega_{k,n} \textbf{m}_{k,n}^{(j)} = \textbf{m}_0 \times \hat{\bm{H}}_k^{(j,j')} \cdot \textbf{m}_{k,n}^{(j')} \, ,
\end{equation}
where the interaction operator $\hat{\bm{H}}_k^{(j,j')}$ can be written in a closed analytical expression that is independent of the thickness coordinate $z$
\begin{equation}
    \begin{split}
        \hat{\bm{H}}_k^{(j,j')} = \omega_\text{B} \delta_{j,j'} \hat{\bm{I}} + \omega_\text{M} \Big[    
         & \lambda_\text{ex}^2 \,k_j^2 \, \delta_{j,j'} \hat{\bm{I}} 
         + \left ( \hat{\textbf{z}} \otimes \hat{\textbf{z}} \right) \delta_{j,j'} \\
         & + \left( \hat{\textbf{k}} \otimes \hat{\textbf{k}}  - \hat{\textbf{z}} \otimes \hat{\textbf{z}} \right) P_{j,j'} \\
         & + \text{i} \left( \hat{\textbf{k}} \otimes \hat{\textbf{z}}  + \hat{\textbf{z}} \otimes \hat{\textbf{k}} \right) Q_{j,j'} \Big] \, .
    \end{split}
\end{equation}
Here, $k_j = \sqrt{k^2 + (j\pi/L)^2}$ is the total wavevector and
\begin{equation} \label{funct} \nonumber
	\begin{split}	
		P_{j,j'} = &\frac{k^2}{k_{j'}^2} \delta_{j,j'} - \frac{k^4}{k_{j}^2 k_{j'}^2} F_{j} \\
        & \times \frac{1}{\left[ (1+\delta_{0,j}) (1+\delta_{0,j'}) \right]^{1/2}} \left( \frac{1+(-1)^{j+j'}}{2} \right) \,,
		\\ 
        \\
		Q_{j,j'} = &\frac{k^2}{k_{j'}^2} \left( \frac{j'^2}{j'^2 -j^2} \frac{2}{k L} - \frac{k^2}{2k_j^2}F_j \right) \\
        &\times \frac{2}{\left[ (1+\delta_{0,j}) (1+\delta_{0,j'}) \right]^{1/2}} \left( \frac{1-(-1)^{j+j'}}{2} \right) \, .
	\end{split}
\end{equation}
are the transformed matrix elements of the dipolar Green's function with
\begin{equation}
F_j = \frac{2}{kL} \left( 1 - (-1)^j e^{-k L}\right) \, .
\end{equation}
To solve for the dispersion relation, \eqref{eq:LLG_cos_basis} is truncated to a sufficiently large number of cos-modes $j=0,1,\dots, j_\text{max}$ and numerically diagonalized afterwards.

\subsection{Analytical Phase Shifter Model: Extraction From The Dispersion}

\begin{figure}[ht]
    \centering
    \includegraphics[width=0.9\linewidth]{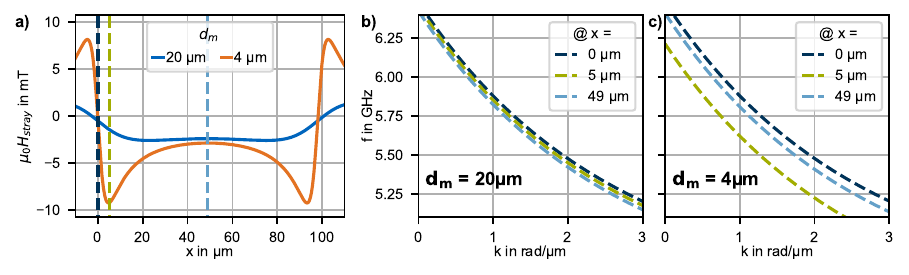}
    \caption{a) The analytically calculated stray field profile along the propagation direction $x$ of the SWs from the input at $x=\SI{0}{\micro\meter}$ to the output at $x=\SI{98}{\micro\meter}$ at the YIG film surface. The analytically calculated $H_\mrm{stray}(x)$ is shown for $\dm=\SI{20}{\micro\meter}$ and $\dm=\SI{4}{\micro\meter}$. At positions $x=[0,5,49]\unit{\micro\meter}$, the SW dispersions at the local effective field $H_\mrm{eff}(x)=H_\mrm{ext}+H_\mrm{stray}(x)$ can be calculated. b) The analytically computed SW dispersions at $\dm=\SI{20}{\micro\meter}$. c) The analytically computed SW dispersions at $\dm=\SI{4}{\micro\meter}$.}
    \label{fig:Hero8_DispersionCurves}
\end{figure}

The data for Fig.\ref{fig:Concept}b-d is extracted from the above-described analytical model for an ideal plan-parallel actuation of the micromagnet.
To emphasize the local change of the dispersion with $H_\mrm{stray}(x)$, we provide the analytically calculated dispersion curves at three propagation distances: Underneath the center of the signal line of the input transducer at $x=\SI{0}{\micro\meter}$, which is the point at which $H_\mrm{stray}$ should not change with $\dm$. At $x=\SI{5}{\micro\meter}$, which is the location of the maximum change in $H_\mrm{stray}$, and at $x=\SI{49}{\micro\meter}$, which is the location of the minimum change in $H_\mrm{stray}$. The extraction is performed as an example for the maximum and minimum tested $\dm$.
The positions are marked via dashed lines in Fig.\ref{fig:Hero8_DispersionCurves}a.
The dispersion curves were calculated with the material parameters provided in Sec.\ref{sec:Fabrication_Methods} and in the above paragraphs of this Section.
The SW dispersion curves at the three positions and at $\dm=\SI{20}{\micro\meter}$ and $\dm=\SI{20}{\micro\meter}$ are shown in Fig.\ref{fig:Hero8_DispersionCurves}b-c, respectively.
For $\dm=\SI{20}{\micro\meter}$, the stray field of the magnet already shifts the local dispersion slightly.
This state is close to the reference phase state, i.e., $\Delta\varphi=\SI{0}{\degree}$, to which all other phases at varying $\dm$ are referenced.

\section{Fabrication and Characterization of the Individual Components}
\label{sec:A_IndividualComponents}

\subsection{Magnonic Chip Before Flip-Chip Process}
\label{ssec:A_MagnonicChip}

The magnonic chip contains the SW transducers, RF feed lines, and DC routing for the MEMS chip. 
The transducers and feed lines were fabricated on the full-film YIG chip using a PMMA multilayer resist patterned by electron-beam lithography (\SI{50}{\kilo\volt} acceleration voltage), followed by lift-off. 
The metallization ($\mathrm{SiO_2}/\mathrm{Cr}/\mathrm{Cu}/\mathrm{Cr}/\mathrm{Au}$ with layer thicknesses of $\SI{10}{\nano\meter}/\SI{10}{\nano\meter}/\SI{1000}{\nano\meter}/\SI{10}{\nano\meter}/\SI{100}{\nano\meter}$) was deposited by electron-beam physical vapor deposition.

\begin{figure}[ht]
    \centering
    \includegraphics[width=0.9\linewidth]{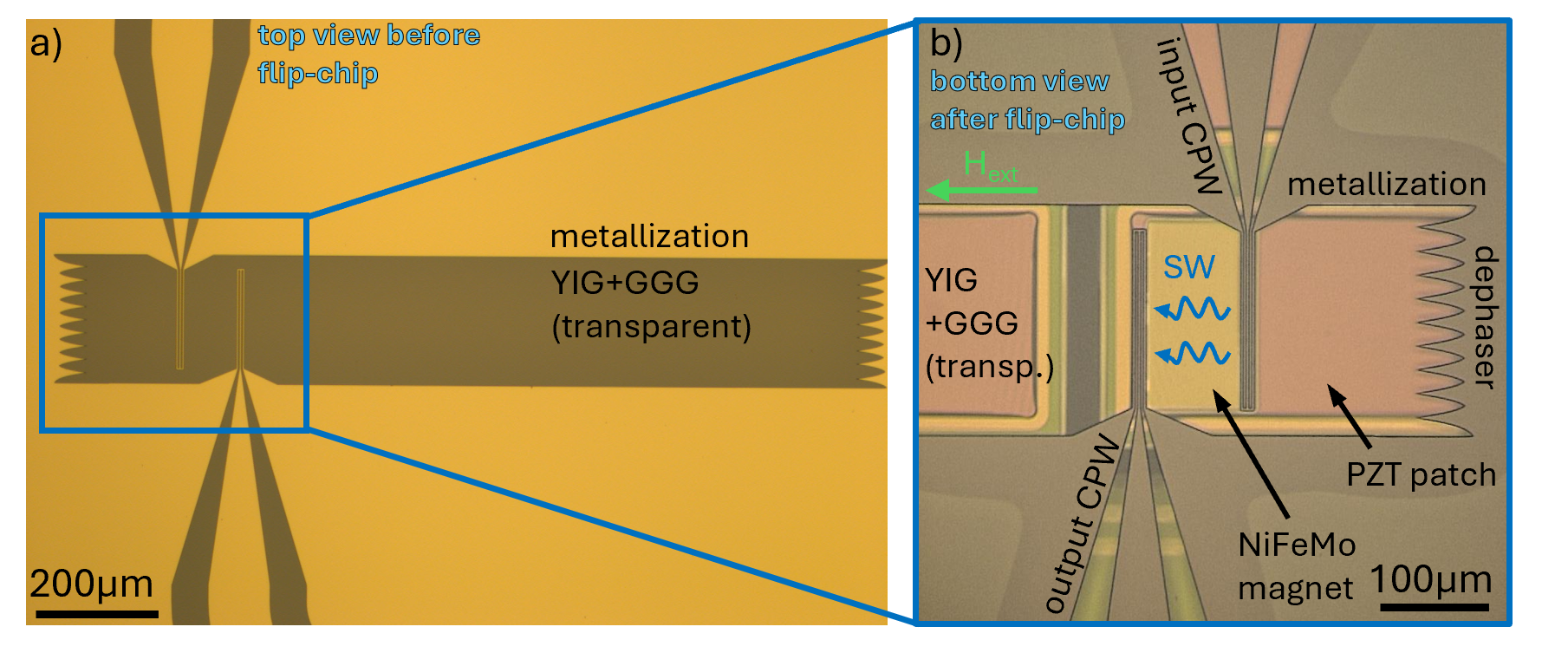}
    \caption{a) Microscope image of the magnonic chip before flip-chip bonding. The transducer periphery is fully metalized to reduce crosstalk and RF losses. The open window allows undisturbed SW transmission in the core phase-shifting area between the CPW transducers, but is reduced (on the left) when the feed lines of a neighboring channel are routed close by. b) A microscope image after flip-chip bonding through the transparent magnonic chip from the underside.}
    \label{fig:TransducerOverview}
\end{figure}

To reduce electromagnetic crosstalk and ensure RF confinement within the CPW feed lines, most of the chip surface is metalized, while dedicated openings around the core phase-shifting region preserve undisturbed SW transmission (Fig.~\ref{fig:TransducerOverview}a). 
Potential SW reflections are further mitigated by dephasing structures at the window edges. 
The openings were designed to maximize the unobstructed SW propagation area while maintaining a minimum separation of twice the CPW width between the inner feed-line edges (signal line width: \SI{100}{\micro\meter}, gap: \SI{60}{\micro\meter}, GGG substrate with $\varepsilon_r=12$~\cite{Connelly_Complex}) to provide a sufficient ground reference and suppress electromagnetic crosstalk between adjacent channels.
The transducers use a ground-signal-ground (GSG) configuration with conductor widths of \SI{2}{\micro\meter}, a gap of \SI{3}{\micro\meter}, and a length of \SI{160}{\micro\meter}. 
The center-to-center spacing between the transducers is \SI{98}{\micro\meter}. 
A microscope image of the SW transducers and the piezoelectric MEMS cantilever carrying the soft magnet, recorded through the transparent YIG chip from the substrate side, is shown in Fig.~\ref{fig:TransducerOverview}b.

\begin{figure}[ht!]
    \centering
    \includegraphics[width=0.9\linewidth]{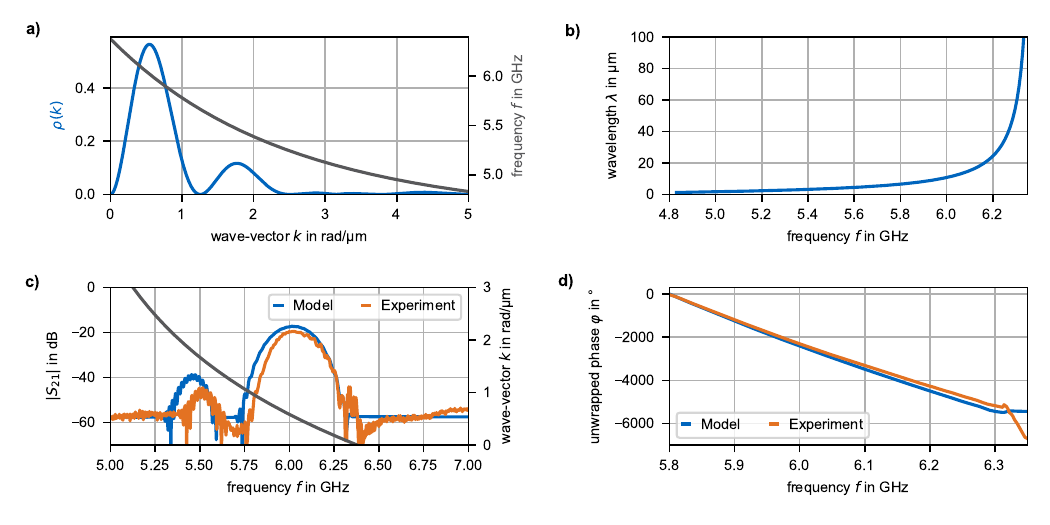}
    \caption{a) Calculated and normalized SW excitation spectrum $\rho(k)$ of the CPW SW transducer together with the analytical dispersion relation, showing optimal excitation around \SI{6}{\giga\hertz}. b) Corresponding SW wavelength derived from the dispersion relation in a). c)–d) Comparison of measured and analytically calculated transmission spectra for magnitude and phase. The measurement was performed at an external field of $\mu_0 H_{\text{ext}}= \SI{147.2}{\milli\tesla}$. The dispersion relation from a) is additionally included in c). For the calculation, an offset field of \SI{1}{\milli\tesla} was used together with $M_\text{eff}= \SI{159.8}{\kilo\ampere\per\meter}$, exchange constant $A_\text{ex}=\SI{3.5}{\pico\joule\per\metre}$, and Gilbert damping $\alpha=2\cdot 10^{-4}$.}
    \label{fig:ModelMeasurement}
\end{figure}

Prior to flip-chip bonding, the magnonic chip was characterized independently. 
Because spin waves are dispersive, the transducer's wavevector excitation spectrum $\rho(k)$ is projected onto the frequency axis through the SW dispersion relation. 
Figure~\ref{fig:ModelMeasurement}a shows the normalized excitation spectrum $\rho(k)$ of the employed CPW SW transducer, calculated from the spatial Fourier transform of the Oersted-field distribution in the transducer cross-section~\cite{Vanderveken_Lumped,Vlaminck_2010}. 
The analytical SW dispersion relation is plotted alongside, indicating a maximum excitation efficiency at approximately \SI{6}{\giga\hertz} or $k_\mrm{SW}\approx\SI{0.6}{\radian\per\micro\meter}$ ($\lambda_\mrm{SW}\approx\SI{10}{\micro\meter}$, see Fig.~\ref{fig:ModelMeasurement}b).
The SW dispersion and transducer excitation spectrum $\rho(k)$ from Fig.~\ref{fig:ModelMeasurement}a were used to calculate the SW transmission spectrum shown in Fig.~\ref{fig:ModelMeasurement}c using the lumped-element model described above. 
Because the dispersion decreases with increasing wavevector (corresponding to shorter SW wavelengths or lower frequencies at fixed $\Hext$ in the BV geometry) and the excitation spectrum peaks near \SI{0.6}{\radian\per\micro\meter}, the transmission maximum occurs around \SI{6}{\giga\hertz}. 
The weaker transmission band near \SI{5.45}{\giga\hertz} originates from the second maximum of the excitation spectrum at $k\approx\SI{1.9}{\radian\per\micro\meter}$. 
In both the model and the experiment, the modulation (ripples) of the transmission response arises from electromagnetic coupling between the transducers. 
For the model calculation, this contribution was set to $Z_{21,\text{em}}=\SI{0.04}{\ohm}$. 
The off-resonant self-impedance measured at the first port was used as the electrical transducer impedance.

Owing to the short SW wavelengths, the model predicts substantial phase accumulation, which is observed experimentally, but is fixed as no MEMS with micromagnets are placed on top of the chip at this stage. 
The resulting phase evolution (Fig.~\ref{fig:ModelMeasurement}d) shows good agreement between model and experiment. 
The same model is used in Supplementary Information Sec.~\ref{ssec:A_Tunability_Analytical} to investigate the tunability of the RF phase shifter beyond \SI{10}{\giga\hertz}.

\subsection{MEMS Chip Fabrication and Characterization}
\label{ssec:A_MEMScharacterization}

\begin{figure}[ht!]
    \centering
    \includegraphics[width=0.8\linewidth]{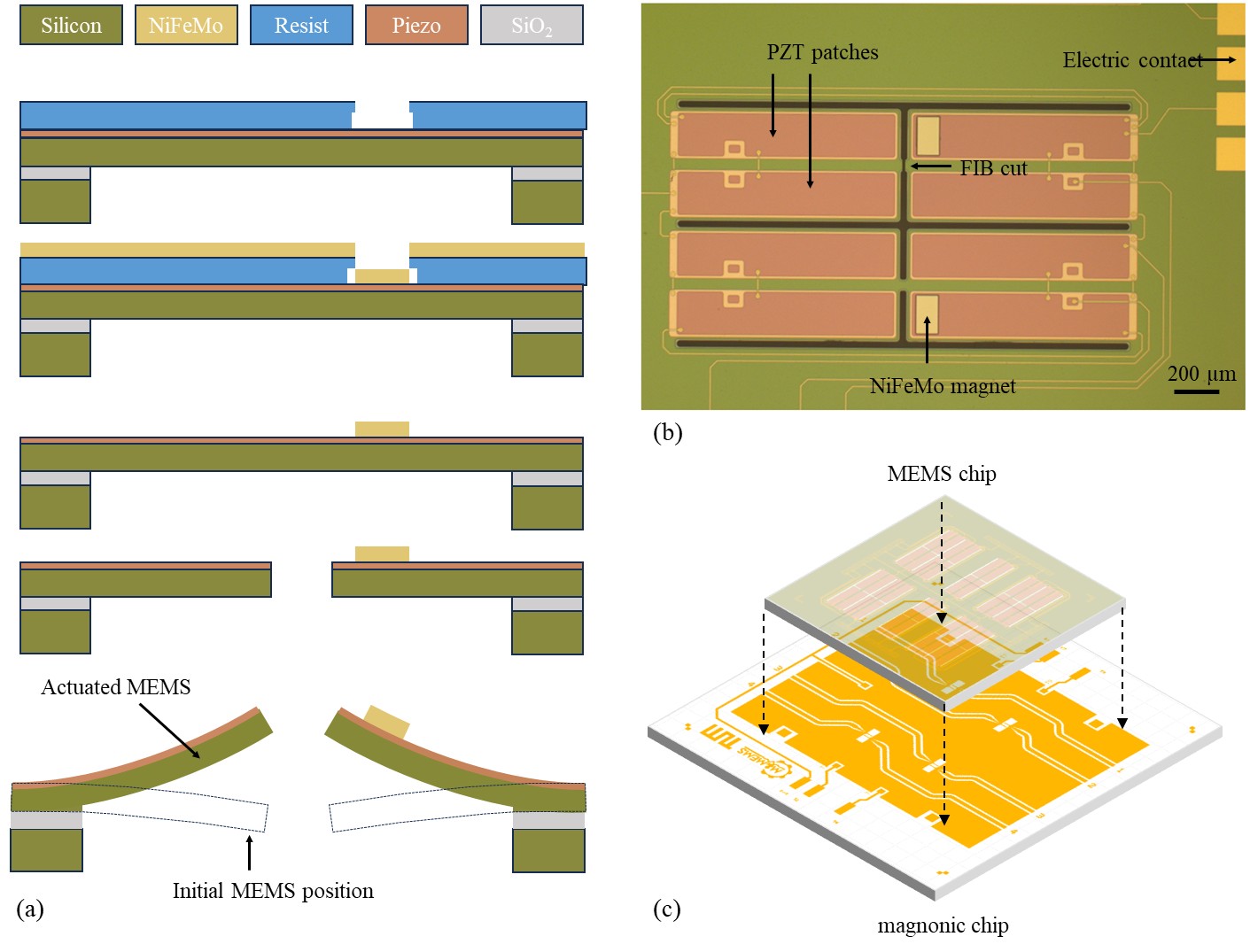}
    \caption{a) Schematic overview of the MEMS chip preparation. After patterning a resist bilayer on the cantilevers via optical lithography, the \SI{695}{\nano\meter} thick Cr/NiFeMo/Cr multilayer magnets are sputter-deposited and structured using lift-off. Subsequently, the silicon joint between the cantilever pair is removed via FIB milling to release the single cantilevers. b) Microscope image of the underside of the MEMS chip highlighting the piezoelectric patches (orange), the NiFeMo micromagnets, and the FIB cut between the cantilever pairs that face each other. c) Sketch of the chip alignment during the flip-chip process showing the full layout of the magnonic chip (\SI{10}{\milli\meter}$\times$\SI{10}{\milli\meter}) and MEMS chip (\SI{6}{\milli\meter}$\times$\SI{6}{\milli\meter}) facing each other.}
    \label{fig:MEMS_fab}
\end{figure}

A schematic overview of the MEMS chip preparation is shown in Fig.~\ref{fig:MEMS_fab}a.
The micromagnets are fabricated on the MEMS cantilever in three steps: (i) a bilayer of a lift-off resist and AZ5214E is structured via optical lithography; (ii) a $\mathrm{Cr}/\mathrm{Ni}_{75}\mathrm{Fe}_{20}\mathrm{Mo}_{5}/\mathrm{Cr}$ multilayer is deposited by sputtering to a total thickness of \SI{695}{\nano\meter}, of which \SI{640}{\nano\meter} is magnetic material; and (iii) the magnets are defined by lift-off of the photoresist mask.
The micromagnets have a saturation magnetization $M_\mrm{S}$ of about \SI{640}{\kilo\ampere\per\meter}.
Afterward, the silicon joint between each pair of cantilevers is removed using focused-ion-beam (FIB) milling to release the two structures and enable large displacements.
An overview microscope image of the underside of two pairs of cantilevers is presented in Fig.~\ref{fig:MEMS_fab}b, showing the MEMS chip with the micromagnets fabricated on the cantilevers.
The micromagnets have lateral dimensions of $\SI{170}{\micro\meter}~\times~\SI{98}{\micro\meter}$, whereas the latter is equivalent to the center-to-center distance $D$ between the transducers on the magnonic chip to achieve maximum phase shift with minimal change of the transmission band.
The magnonic chip and the MEMS chip are then connected to each other by flip-chip bonding (Figure~\ref{fig:MEMS_fab}c). 
Therefore, a set of gold stud bumps is formed on the magnonic chip using a wire bonder. 
The two chips are then aligned and bonded via thermocompression using a Finetech Fineplacer Pico~1 system, which offers a nominal lateral alignment accuracy of \SI{5}{\micro\meter}. 
During thermocompression, the chips are heated to \SI{360}{\celsius} for \SI{150}{\second}, while a force of \SI{0.033}{\newton} per interconnect is applied.
Four large bond pads at the corners of the magnonic chips are used solely for mounting, while several smaller bond pads provide a DC electrical connection between the two chips.
The DC supply for the MEMS chip is therefore provided via the magnonic chip bonded to the PCB.

\subsubsection*{MEMS Characterization in the Hybrid Device}

\begin{figure}[ht!]
    \centering
    \includegraphics[width=0.9\linewidth]{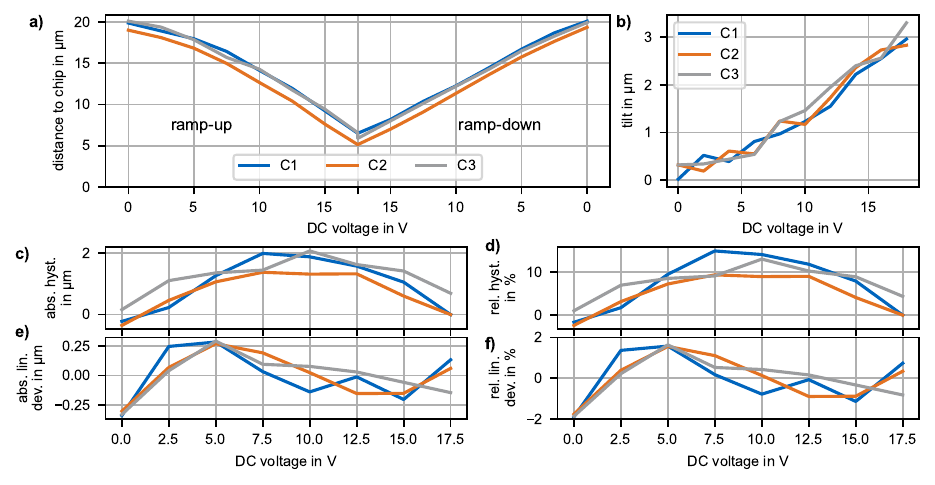}
    \caption{Displacement characterization of the piezoelectric cantilevers C1,~C2,~C3 after \num{750} conditioning cycles. a) While the cantilevers show a slightly quadratic displacement when ramping up the voltage, it is possible to obtain an almost linear displacement when ramping down the applied voltage. b) The voltage-dependent tilt of the cantilever along the SW propagation direction. The tilt is calculated at the positions of the input and output transducers on the magnonic chip. The absolute c) and relative d) hysteresis of the cantilevers between voltage ramp-up and ramp-down. The absolute e) and relative f) linearity deviations for the cantilevers for ramping down the applied voltage from the pre-defined state.}
    \label{fig:MEMS_Characterization}
\end{figure}

For conditioning of the piezoelectric MEMS cantilevers, \num{750} voltage ramps between \SI{0}{\volt} and \SI{20}{\volt} (and back) were applied in steps of \SI{0.1}{\volt}, without allowing contact between the cantilever and the opposing chip surface. 
A confocal 3D laser-scanning microscope with a vertical resolution of up to \SI{10}{\nano\meter} was used to determine the distance between the cantilever (top or backside) and the magnonic chip surface at positions above the input and output transducers. 
Figure~\ref{fig:MEMS_Characterization}a shows the distance between the underside of the micromagnet and the magnonic chip surface $\dm$ as a function of applied DC voltage for the three investigated cantilevers.

An almost linear displacement–voltage relation can be obtained by pre-biasing the cantilever at the maximum voltage and subsequently ramping down. 
Because the cantilevers are clamped on one side, the free end is tilted along the SW propagation direction, and the attached micromagnet follows the same tilt. 
The resulting height offset between the input and output transducer positions is shown in Fig.~\ref{fig:MEMS_Characterization}b, with a maximum tilt-induced offset of approximately \SI{3.5}{\micro\meter} at the highest applied voltage. 
After conditioning, the cantilevers exhibit a typical absolute and relative hysteresis of \SI{2}{\micro\meter} and \SI{15}{\percent}, respectively (Fig.~\ref{fig:MEMS_Characterization}c–d), where the relative hysteresis is normalized to the full displacement range.
The absolute and relative linearity deviation during voltage ramp-down from the pre-biased state at \SI{20}{\volt} to \SI{0}{\volt} are shown in Fig.~\ref{fig:MEMS_Characterization}e–f. 
The deviations do not exceed \SI{350}{\nano\meter} and \SI{2}{\percent}, respectively, with the relative value normalized to the maximum of $\dm$.
All data presented in this work were recorded in the voltage ramp-up mode and using a look-up table for the voltage-to-distance translation based on a fitted quadratic function per cantilever.
The fitted quadratic functions show degrees of definiteness of better than \num{0.998}, i.e., less than \SI{2}{\percent} deviation, which is in the same order of magnitude as the intrinsic non-linearity of the cantilevers when ramping down the applied DC voltage.

The hybrid chip was mounted on the carrier PCB using rubber cement, ensuring a flat placement and minimizing the adhesive layer thickness to facilitate subsequent bonding. 
Electrical interconnection to the PCB was realized using either gold ball bonds or aluminum wedge bonds with a wire diameter of \SI{25}{\micro\meter} (see also Fig.~\ref{fig:Concept}h in the main text). 

\subsection{RF Carrier Board}
\label{ssec:A_CarrierBoard}

The RF carrier PCB was primarily designed as a practical mounting platform for the hybrid chip, enabling DC biasing of the MEMS actuators and facilitating device operation. 
The design prioritizes interface/physical compatibility between the hybrid chip and the PCB, while impedance matching was intentionally not performed at this stage.
Further reductions in insertion loss and footprint are possible through dedicated layout optimization in future realizations.

The carrier PCB is a four-layer FR4 board with a total thickness of \SI{1.55}{\milli\meter}. 
In the center of the board, a short section of four CPW traces is implemented, which can be wire-bonded for RF characterization and later removed by milling to accommodate the hybrid chip. 
The CPW traces are designed for a \SI{50}{\ohm} characteristic impedance (line width: \SI{254}{\micro\meter}, gap: \SI{200}{\micro\meter}, dielectric thickness above the RF ground plane: \SI{140}{\micro\meter}, $\varepsilon_r\approx4.3$; all nominal values). 
The CPW trace lengths differ by up to \SI{2.5}{\milli\meter}, corresponding to a delay of approximately \SI{4}{\pico\second} per side, assuming an effective dielectric constant of \num{2.5} and a velocity factor of about \SI{63}{\percent}~\cite{Ghione1987}. 
This delay variation is negligible compared to the SW propagation delay of approximately \SI{28}{\nano\second} at $f_\mrm{c}=\SI{6.1}{\giga\hertz}$ (see also Supplementary Information Sec.~\ref{ssec:A_GroupDelay}).
Right-angled surface-mount SMP connectors with a maximum voltage-standing-wave ratio (VSWR) of \num{1.10} (corresponding to a return loss of \SI{-26.8}{\decibel} up to \SI{8}{\giga\hertz}) were used to provide four RF access lines for the input and output sides of the hybrid chip.

For RF characterization, the central CPW section was wire-bonded and measured using a Rohde~\&~Schwarz ZNB20 vector network analyzer (VNA). 
Calibration was performed using a standard TOSM two-port calibration to define the reference plane at the ends of the \SI{3.5}{\milli\meter} coaxial cables. 
Transition to the SMP connectors was realized using \SI{30.5}{\centi\meter} adapter cables (TFLEX 405) and de-embedded according to the IEEE370-2020 standard~\cite{IEEE370}.
The four CPW traces on the carrier PCB exhibit a return loss of \num{4.7}$\pm$\SI{2}{\decibel} and an insertion loss of \num{4.7}$\pm$\SI{1.5}{\decibel} at \SI{6.1}{\giga\hertz}. 
Up to \SI{3.8}{\giga\hertz}, the insertion loss remains below \SI{3}{\decibel}, and up to \SI{9.9}{\giga\hertz} below \SI{10}{\decibel}  for all CPW traces.

Crosstalk between CPW traces shows a pronounced frequency dependence. 
Up to approximately \SI{5}{\giga\hertz}, coupling between neighboring RF traces lies in the range of \SIrange{-30}{-20}{\decibel}, while next-nearest neighbors (1–3 and 2–4) exhibit on average about \SI{10}{\decibel} lower coupling. 
Crosstalk between the outermost channels (1–4) is approximately \SI{15}{\decibel} lower than that of adjacent traces. 
Above \SI{5}{\giga\hertz}, this trend becomes less pronounced as the guided wavelength on the CPW structures (\SI{30}{\milli\meter} at \SI{6}{\giga\hertz} for the given FR4 geometry~\cite{Ghione1987}) approaches the physical trace length (\SI{20}{\milli\meter} per side), leading to resonances in the transmission lines, bond wires, and bonded CPW section. 
Crosstalk measurements were performed with all ports open except for the two actively measured ports.

\section{Post-Processing and Supplementary Measurements}
\label{sec:A_PostProcessing}

\subsection{Time-Domain Gating}
\label{ssec:A_TimeDomainGating}
\begin{figure}[ht]
    \centering
    \includegraphics[width=0.9\linewidth]{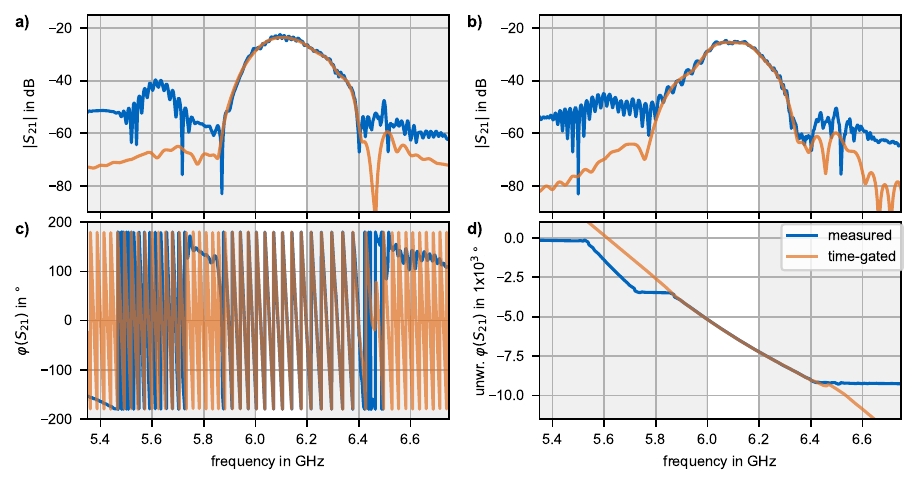}
    \caption{a)-b) Comparison of the measured and time-gated $|S_\mrm{21}|$ at $\dm=\SI{20.7}{\micro\meter}$ and $\dm=\SI{4}{\micro\meter}$, respectively. c) While the rapid phase accumulation with frequency is challenging to track in the wrapped $\varphi(S_{21})$, unwrapping facilitates the picture in d). Time gating allows for a strong reduction of additional effects, such as intermodulation of the SW signal with electromagnetic crosstalk or multi-path SW propagation~\cite{Kohl_ModellingSW,Davidkova2025}. By applying time-gating around the desired main SW contribution, it is possible to extract the idealized device behavior without altering its principal characteristics.}
    \label{fig:RF1_ComparisonTimeGating}
\end{figure}

To obtain a more detailed view of the ideal SW characteristics, we apply a time-gating procedure similar to that reported in~\cite{Davidkova2025,Kohl_ModellingSW}, with the additional use of a frequency-domain windowing filter to further suppress out-of-band signal contributions.
Owing to the significantly lower group velocity of SWs compared to microwave signals at the same frequency, time-gating enables isolation of the SW-related signal contribution that would otherwise be affected by electromagnetic crosstalk or multipath SW propagation.
For frequency-domain windowing, a Tuckey filter was used, i.e., a tapered cosine window with a flat passband~\cite {Tuckey1967}, ensuring that phase and amplitude information within the \SI{200}{\mega\hertz} analysis bandwidth remain unaffected.
The measured and time-gated $|S_{21}|$ is shown in Fig.~\ref{fig:RF1_ComparisonTimeGating}a-b for $\dm=\SI{20.7}{\micro\meter}$ and $\dm=\SI{4}{\micro\meter}$, respectively. 
The highlighted region denotes a \SI{200}{\mega\hertz} bandwidth around $f_\mrm{c}=\SI{6.1}{\giga\hertz}$.
The span is chosen to exemplify the effect of time-gating, with the second excitation maximum of the transducer around \SI{5.6}{\giga\hertz}.
Additionally, the measured and time-gated $\varphi(S_{21})$ at $\dm=\SI{20.7}{\micro\meter}$ is shown in Fig.~\ref{fig:RF1_ComparisonTimeGating}c, while the corresponding unwrapped phase is shown in Fig.~\ref{fig:RF1_ComparisonTimeGating}d.
Time-gating suppresses additional contributions such as intermodulation between the SW signal and electromagnetic crosstalk, as well as multipath SW propagation between input and output transducers~\cite{Davidkova2025,Kohl_ModellingSW}.
This yields an idealized representation of amplitude and phase within the relevant SW transmission band, enabling more direct interpretation of device behavior and supporting further optimization.

\subsection{Scattering Parameters and Phase Extraction}
\label{ssec:A_RF_Sparams_Phase}

\begin{figure}[h]
    \centering
    \includegraphics[width=0.9\linewidth]{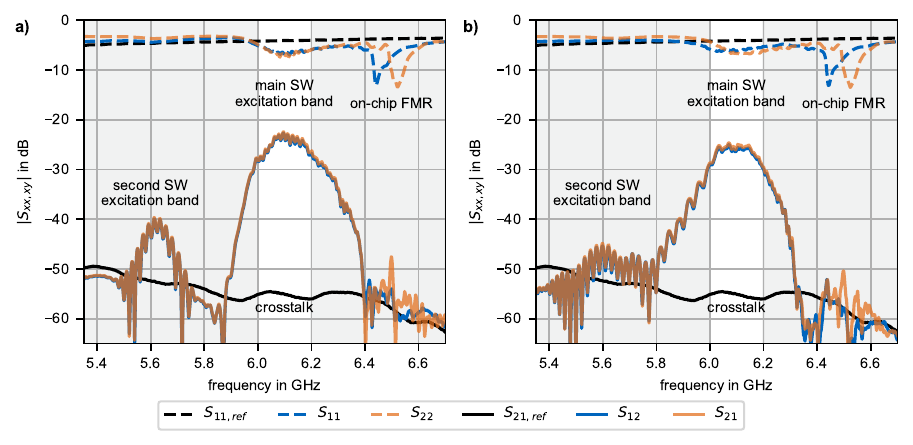}
    \caption{Full 2-port S-parameters at a) $\dm=\SI{20.7}{\micro\meter}$  and b) $\dm=\SI{4}{\micro\meter}$ at an external field of $\mu_0\Hext=\SI{152}{\milli\tesla}$. The S-parameters show a nearly reciprocal behavior except for the FMR excitation around \SI{6.5}{\giga\hertz}. The difference of \SI{0.08}{\giga\hertz} is caused by different RF CPW trace lengths on the magnonic chip on the input/output sides, which experience locally different effective magnetic bias fields $\mu_0\Heff$ of about \SI{2.8}{\milli\tesla}.}
    \label{fig:RF1_2Port_Comparison}
\end{figure}

A complete two-port S-parameter characterization in magnitude is shown in Fig.~\ref{fig:RF1_2Port_Comparison}a at $\dm=\SI{20.7}{\micro\meter}$ and in Fig.~\ref{fig:RF1_2Port_Comparison}b at $\dm=\SI{4}{\micro\meter}$, measured at $\mu_0\Hext=\SI{152}{\milli\tesla}$ and a center frequency of $f_\mrm{c}=\SI{6.1}{\giga\hertz}$.
The S-parameters exhibit approximately reciprocal behavior, with slight deviations near the ferromagnetic resonance (FMR) around \SI{6.5}{\giga\hertz}. 
These deviations are attributed to different CPW feedline lengths on the input and output sides of the magnonic chip, which lead to locally varying effective magnetic bias fields $\Heff$ near the chip edges.
The observed frequency offset of \SI{0.08}{\giga\hertz} between the FMR peaks in $S_{11}$ and $S_{22}$ corresponds to a difference of approximately \SI{2.8}{\milli\tesla} in the local effective field $\mu_0\Heff$.
The reflection parameters further indicate a minimum reflected power approximately \SI{400}{\mega\hertz} above $f_\mrm{c}$ at wave vectors approaching the FMR regime. 
This is attributed to the excitation and detection of on-chip FMR beneath the wide CPW feed lines outside the defined SW transmission channel.
Since this contribution is irrelevant to the device's functionality, it could be mitigated by locally removing the YIG film outside the transmission region and by restricting propagation to patterned YIG islands that define the spin-wave channel.

Using a reference measurement at $\mu_0\Hext=\SI{103}{\milli\tesla}$, the electromagnetic system behavior can be compared with and without the SW transmission channel acting as a load.
For example, the approximately \SI{3}{\decibel} RL at \SI{6.1}{\giga\hertz} in the reference measurement indicates that about half of the input power is dissipated in the CPW traces due to ohmic and minor dielectric losses.
While the ideal expected RL would approach \SI{0}{\decibel} due to the SW transducers being shorted at their ends, approximately \SI{50}{\percent} of the input power is dissipated, and the remaining \SI{50}{\percent} is reflected back to the VNA.
When tuning the SW system to $f_\mrm{c}=\SI{6.1}{\giga\hertz}$, the RL decreases to about \SI{6}{\decibel}, indicating that approximately half of the available input power at the SW transducer is transferred into the SW transmission channel.
Consequently, only about one quarter of the input power is reflected back to the port~\cite{Kohl_identification}.
The variation of input and output RL at \SI{6.1}{\giga\hertz} over the full cantilever displacement range is approximately \SI{0.75}{\decibel}.
The change in RL between the electromagnetic reference state and the tuned operational point can be interpreted as an effective matching contribution arising from power transduction into the SW channel.

\begin{figure}
    \centering
    \includegraphics[width=0.9\linewidth]{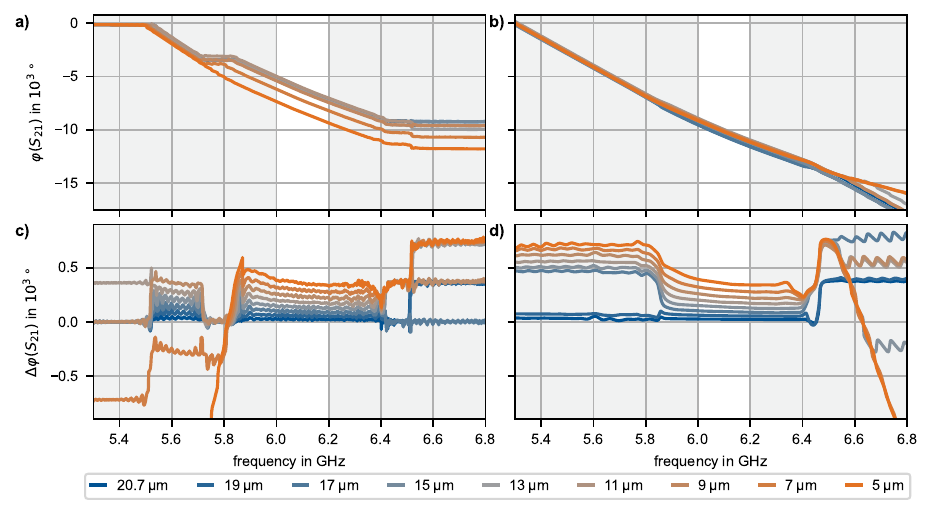}
    \caption{a) The phase states calculated from the unwrapped $\varphi(S_\mrm{21})$ at a constant $\mu_0\Hext=\SI{152}{\milli\tesla}$, where large spin-wave phase accumulation dominates and intermodulation with electromagnetic crosstalk introduces integer-multiple 360° offsets. b) Time-gating the measured signal suppresses these offsets. The resulting shift $\Delta\varphi(S_{21})$ after subtracting the reference phase state ($\dm=\SI{20.7}{\micro\meter}$) and correcting the remaining $n\times\SI{360}{\degree}$ offsets at $f_\mrm{c}=\SI{6.1}{\giga\hertz}$ for the measured c) and time-gated d) phase information.}
    \label{fig:RF1_PhaseCalc}
\end{figure}

The unwrapped phase of the measured S-parameters is used to determine the phase states for all cantilever positions at constant $\mu_0\Hext=\SI{152}{\milli\tesla}$, as shown in Fig.~\ref{fig:RF1_PhaseCalc}a.
Intermodulation between the SW signal and electromagnetic crosstalk affects the phase unwrapping procedure and introduces occasional offsets of integer multiples of \SI{360}{\degree}.
This effect can be mitigated by applying time-gating prior to phase extraction, as shown in Fig.~\ref{fig:RF1_PhaseCalc}b.
The reference phase is defined at $\dm=\SI{20.7}{\micro\meter}$, and is subtracted from all subsequent measurements at varying cantilever positions.
Residual $n\times\SI{360}{\degree}$ offsets are corrected at the center frequency $f_\mrm{c}=\SI{6.1}{\giga\hertz}$, resulting in the data shown in Fig.~\ref{fig:RF1_PhaseCalc}c.
The time-gated data is processed analogously and shown in Fig.~\ref{fig:RF1_PhaseCalc}d.
Distortions in the second excitation band between \SIrange{5.5}{5.75}{\giga\hertz} originate from residual intermodulation between the SW signal and electromagnetic crosstalk, which accumulates phase but does not satisfy the $\pm\SI{180}{\degree}$ condition required for stable phase unwrapping.
The offset correction is required only due to the relatively broadband characterization of the magnonic phase shifter over a \SI{4.2}{\giga\hertz} span.
In practical operation, this effect is naturally suppressed because the usable bandwidth is limited to the device's operating band.

\subsection{Variation of Center Frequency, Insertion/Return Loss, and Bandwidth with Cantilever Actuation}
\label{ssec:A_Variation_fc_IL_BW}

\begin{figure}[t!]
    \centering
    \includegraphics[width=0.9\linewidth]{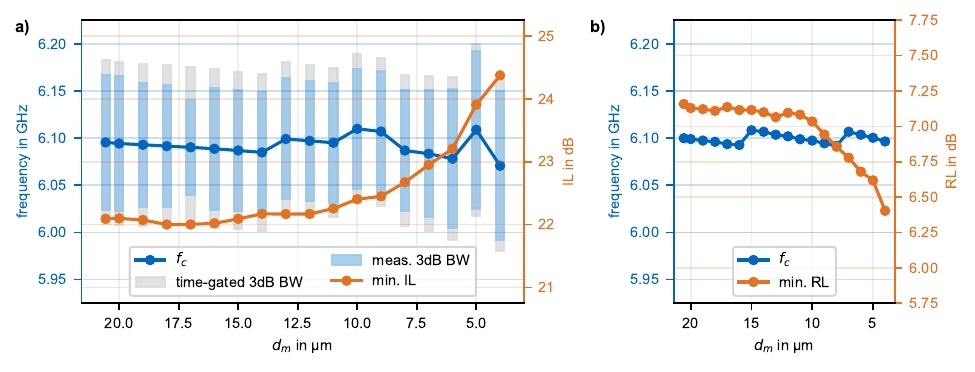}
    \caption{a) Variation of the center frequency $f_\mrm{c}$, the $\BWthreeDB$, and the IL at $f_\mrm{c}$ versus the characterized magnet distances to the magnonic chip $\dm$. The center frequency shows a maximum shift of \SI{24.8}{\mega\hertz} at $\dm=\SI{4}{\micro\meter}$ distance, while the IL is increased by about \SI{2.5}{\decibel} compared to $\dm=\SI{20.7}{\micro\meter}$. The mean measured and time-gated $\BWthreeDB$ is \SI{134}{\mega\hertz} and \SI{167}{\mega\hertz}, respectively. b) Variation of $f_\mrm{c}$ extracted from the minimum RL with varying magnet distances $\dm$. The RL changes at most \SI{0.75}{\decibel} across all cantilever positions.}
    \label{fig:RF1_Fcenter_BW_Mag_Combi}
\end{figure}

The residual variation of the center frequency $f_\mrm{c}$, the minimum IL, and the $\BWthreeDB$ with varying cantilever positions are presented in Fig.~\ref{fig:RF1_Fcenter_BW_Mag_Combi}a.
The external magnetic bias field $\Hext$ was chosen such that $f_\mrm{c}=\SI{6.1}{\giga\hertz}$ at the highest (unbiased) cantilever position.
By sweeping through the cantilever positions, $f_\mrm{c}$ shifts by a maximum of \SI{24.8}{\mega\hertz}.
This behavior is reproduced in micromagnetic simulations and attributed primarily to the cantilever tilt and the resulting locally slightly different $\Heff$ at the input and output transducers (see also Supplementary Information Sec.~\ref{ssec:A_ComparisonMuMax}).
In addition, cantilever actuation shifts the positions of ripples in the transmitted SW response, leading to a slightly altered detection of the minimum IL in frequency when using raw data (no averaging or time-gating).
At the same time, the IL increases by approximately \SI{2.5}{\decibel} at $\dm=\SI{4}{\micro\meter}$.
The measured and time-gated $\BWthreeDB$ are plotted around the detected $f_\mrm{c}$ in blue and gray, respectively.
The mean measured $\BWthreeDB$ is \SI{134}{\mega\hertz}, while time-gating yields $\BWthreeDB=\SI{167}{\mega\hertz}$ due to suppression of ripples arising from intermodulation between electromagnetic crosstalk and multipath SW propagation.
In Fig.~\ref{fig:RF1_Fcenter_BW_Mag_Combi}b, a similar trend is observed for the forward RL, indicating a slight variation of $f_\mrm{c}$ with cantilever actuation.
The minimum detected RL at $f_\mrm{c}$ changes by approximately \SI{0.75}{\decibel}.

\subsection{Extraction of Absolute/Root-Mean-Square Phase and Amplitude Error, and System Linearity \& Sensitivity}
\label{ssec:A_PhaseErrors}

\begin{figure}
    \centering
    \includegraphics[width=0.9\linewidth]{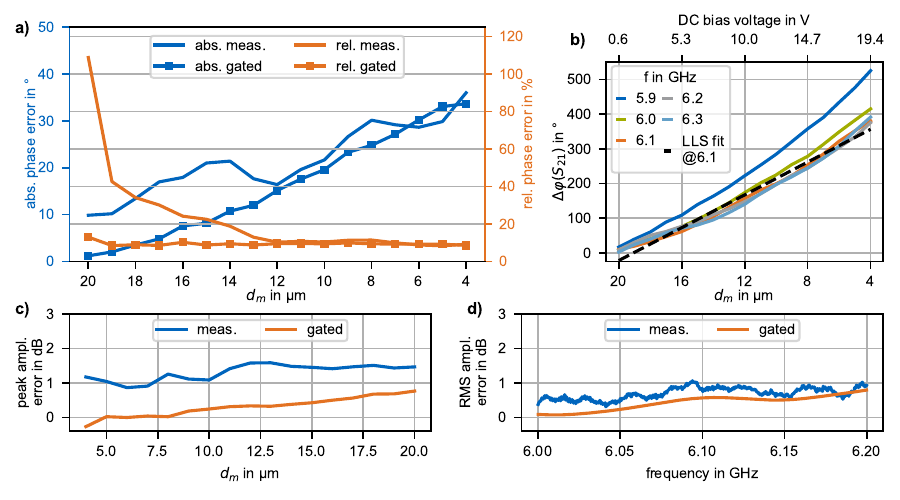}
    \caption{Summary of practical device characteristics. a) The peak absolute phase error for the measured and time-gated signal (left y-axis) together with the relative phase error of both signals (right y-axis), which is normalized to the expected phase shift at each value of $\dm$. b) The phase-voltage-sensitivity around $f_\mrm{c}=\SI{6.1}{\giga\hertz}$ is \SI{20.3}{\degree\per\volt}. Based on this, the phase-voltage-linearity deviation is calculated by fitting a linear-least-squares function to the graph at $f_\mrm{c}=\SI{6.1}{\giga\hertz}$ (black dashed line), and we obtain \SI{32.1}{\degree} in absolute value or \SI{8.5}{\percent} normalized to the full cantilever displacement. c) The peak amplitude error across the varying cantilever positions $d_m$ and d) the RMS amplitude error in a span from \SIrange{6}{6.2}{\giga\hertz} are in the range of \SI{1}{\decibel}. The amplitude errors in the time-gated signals are slightly reduced.}
    \label{fig:RF1_TruePhaseChange}
\end{figure}

The absolute and relative phase errors, together with the root-mean-square (RMS) phase error, are important metrics for evaluating the functionality of RF phase shifters across the operating bandwidth.
In Fig.~\ref{fig:RF1_TruePhaseChange}a, the peak absolute and relative phase errors for both measured and time-gated data are shown as a function of cantilever position.
The peak absolute error is defined as the maximum deviation of the phase within the operational band from the expected phase shift.
The expected phase value is defined as the mean phase shift within $\pm\SI{100}{\mega\hertz}$ around $f_\mrm{c}$ for each cantilever position.
The peak absolute error increases with increasing phase shift $\Delta\varphi$, or equivalently decreasing $\dm$, and represents a worst-case estimate of the phase deviation.
It includes contributions from the dispersive spin-wave behavior with swept frequency at constant $\Hext$ as well as phase ripple induced by intermodulation between electromagnetic crosstalk and multi-path propagating spin waves~\cite{Kohl_ModellingSW,Davidkova2025}.
The relative phase error is normalized to the expected phase shift and remains at approximately \SI{10}{\percent} for both measured and time-gated data.

For calculating the Root-Mean-Square (RMS) phase error $\delta\varphi_\mrm{RMS}(f)$, as presented in Fig.~\ref{fig:RF1_PhaseChange_combi}f in the main part, we applied the following calculation for the $N=18$ evaluated phase states.
First, the phase error 
\begin{equation}
    \delta\varphi_i(f) = \Delta\varphi(S_{21})(f)-\Delta\varphi_\mrm{i,ideal}
\end{equation}
is calculated with respect to the ideal or expected constant phase shift $\Delta\varphi_\mrm{i,ideal}$ for all measurements $i=1,\ldots,N$.
In this work, we defined $\Delta\varphi_\mrm{i,ideal}$ as the mean phase change of the time-gated signal at each $\dm$, instead of using theoretically defined/expected values, which would be the typical approach for digital/discrete conventional RF phase shifters.
Second, the average phase error 
\begin{equation}
    \label{eq:Mean_PhaseError}
    \overline{\delta\varphi(f)} = \frac{1}{N}\sum^N_i \delta\varphi_i(f)
\end{equation}
is treated as the new or corrected reference state and used to correct the measured phase states, which means that all $N$ states are subtracted by $\overline{\delta\varphi(f)}$ to obtain the corrected phase state diagram shown in Fig.~\ref{fig:Concept}e in the main part.
In this diagram, states around half the maximum phase shift appear flatter, while the previously (ideal) \SI{0}{\degree} state is no longer constant, because subtracting $\overline{\delta\varphi(f)}$ corrects the average slope of all $N$ phase states.
Last, the resulting RMS phase error
\begin{equation}
    \label{eq:RMS_PhaseError}
    \delta\varphi_\mrm{RMS}(f) = \sqrt{\frac{1}{N}\sum^N_i (\delta\varphi_i(f)-\overline{\delta\varphi(f)})^2}
\end{equation}
is calculated with reference to the average phase error $\overline{\delta\varphi(f)}$.
It is common to calculate the RMS phase error relative to the average phase error rather than the ideal/expected phase states, and provide this value in data sheets.
The RMS phase error $\delta\varphi_\mrm{RMS}(f)$ is calculated under the assumption that the $N=18$ measured phase states would represent all available phase states of the presented magnonic phase shifter.

In a similar manner, the peak and RMS amplitude error is calculated: Subtracting the average amplitude
\begin{equation}
    \overline{|S_{21}(f)|} = \frac{1}{N}\sum_i^N|S_{21}(f)|
\end{equation}
from the $N$ measurements, the amplitude errors
\begin{equation}
    \delta|S_{21,i}(f)| = |S_{21,i}(f)|-\overline{|S_{21}(f)|}
\end{equation}
are calculated.
The peak amplitude error $\delta|S_{21,i}(f)|_\mrm{peak} = \max(\delta|S_{21,i}(f)|)$ at each $\dm$ is shown in Fig.~\ref{fig:RF1_TruePhaseChange}c for the measured and time-gated signals.
The peak amplitude error $\delta|S_{21,i}(f)|_\mrm{peak}$ serves as a worst-case estimation of the amplitude variation in the operational band of the RF phase shifter.
The RMS amplitude error 
\begin{equation}
    \label{eq:RMS_AmplitudeError}
    \delta|S_{21}(f)|_\mrm{RMS} = \sqrt{\frac{1}{N}\sum^N_i \delta |S_{21,i}(f)|^2}
\end{equation}
is calculated with reference to the average amplitude error $\overline{|S_{21}(f)|}$ and shown in Fig.~\ref{fig:RF1_TruePhaseChange}d.
The measured and time-gated $\delta|S_{21}(f)|_\mrm{RMS}$ are below \SI{1.1}{\decibel} in a \SI{200}{\mega\hertz} span around $f_\mrm{c}$.

Convenient control of signals in practical applications requires sufficiently linear device response, as well as a quantitative understanding of the associated sensitivities and linearity deviations.
The distance-phase sensitivity and voltage-phase sensitivity, together with the corresponding linearity of the phase change $\Delta\varphi(S_{21})$, are extracted from the data in Fig.~\ref{fig:RF1_TruePhaseChange}b, which shows $\Delta\varphi(S_{21})$ as a function of $\dm$ and applied DC voltage on the lower and upper x-axes, respectively.
For the extraction, a linear least-squares (LLS) fit is applied to the voltage- and distance-dependent phase response at $f_\mrm{c}=\SI{6.1}{\giga\hertz}$, as indicated by the dashed line in Fig.~\ref{fig:RF1_TruePhaseChange}b.
During characterization, hysteresis effects of the piezoelectric cantilevers are accounted for by either using a look-up table during voltage ramp-up or by operating on a voltage ramp-down sequence from a predefined state, enabling an almost linear actuation with applied voltage (see Supplementary Information Sec.~\ref{ssec:A_MEMScharacterization}).
The magnonic phase shifter exhibits a voltage-phase sensitivity of \SI{20.3}{\degree\per\volt}, corresponding to a distance-phase sensitivity of \SI{23.7}{\degree\per\micro\meter}.
The absolute and relative linearity deviation of the phase response at equidistant $\dm$ is \SI{32.1}{\degree} and \SI{8.5}{\percent}, respectively, where these values include the effects of residual piezoelectric nonlinearity and hysteresis.
The relative linearity deviation is normalized to the full travel range of the MEMS cantilever.

\subsection{Group Delay}
\label{ssec:A_GroupDelay}

\begin{figure}[H]
    \centering
    \includegraphics[width=0.9\linewidth]{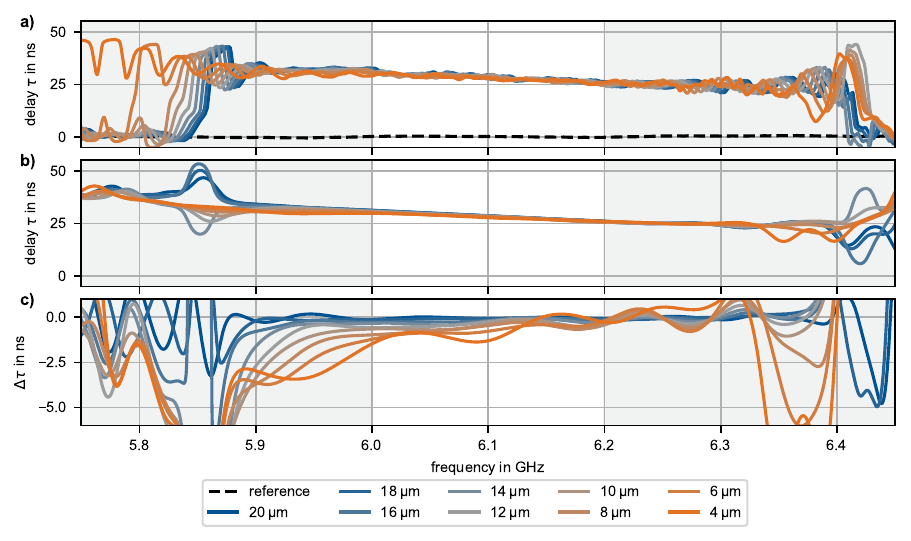}
    \caption{Measured a) and time-gated b) group delay $\tau_\mrm{g}$ after applying a delay aperture $T_\mrm{a}=\SI{10}{\mega\hertz}$ to the signals. The measured data is impaired by the phase ripples due to crosstalk intermodulation and multi-path propagation. c) Delay change $\Delta\tau$ with respect to the cantilever position at maximum distance to the magnonic chip. In a \SI{200}{\mega\hertz} span around $f_\mrm{c}=\SI{6.1}{\giga\hertz}$, $\Delta\tau$ stays within $\pm\SI{1.5}{\nano\second}$ with respect to the measurement at $\dm=\SI{20.7}{\micro\meter}$.}
    \label{fig:RF1_Delay}
\end{figure}

Another way to quantify phase linearity in frequency-dependent RF devices is to compute the group delay $\tau_\mrm{g}(\omega)$, defined as the negative derivative of the phase with respect to angular frequency $\omega$
\begin{equation}
    \label{eq:DelayChange}
    \tau_\mrm{g}(\omega) = -\frac{\partial\varphi(\omega)}{\partial\omega} \approx -\frac{\Delta\varphi(\omega)}{\Delta\omega} = \frac{\varphi(\omega_i)-\varphi(\omega_{i-1})}{\omega_i-\omega_{i-1}}.
\end{equation}
To avoid overemphasizing deviations from the ideal linear phase due to phase noise and the VNA's measurement uncertainty, a delay aperture $T_\mrm{a}$ is commonly introduced.
This aperture computes the finite difference between $\omega_i$ and $\omega_{i-m}$ with $m>1$, effectively smoothing ripples.
The aperture width $T_\mrm{a}$ depends on the measurement configuration and was chosen according to~\cite{Ostwald1998GroupAP} as \SI{10}{\mega\hertz}, corresponding to a delay uncertainty of approximately \SI{100}{\pico\second}.
This uncertainty is about 300 times smaller than the measured absolute SW-induced delay.
The group delay $\tau_\mrm{g}$ was evaluated using $T_\mrm{a}=\SI{10}{\mega\hertz}$ for both measured and time-gated signals, as shown in Fig.~\ref{fig:RF1_Delay}a-b.
Due to the dispersive nature of SWs with swept frequency at constant bias field, the group delay varies with frequency across the accessible SW transmission band.
The measured delay contains pronounced ripples arising from phase modulation effects associated with electromagnetic crosstalk and spin-wave interference~\cite{Davidkova2025,Kohl_ModellingSW}.
After time-gating, the group delay variation is reduced and remains within approximately \SI{1.5}{\nano\second} over the \SIrange{6}{6.2}{\giga\hertz} band.
The change in group delay $\Delta\tau$ relative to the reference configuration at $\dm=\SI{20.7}{\micro\meter}$ is shown in Fig.~\ref{fig:RF1_Delay}c.
While group-delay variation is typically undesirable for phase shifter operation, the observed dispersive regime between \SI{5.9}{\giga\hertz} and \SI{6}{\giga\hertz} indicates that the same device concept can also operate as a tunable delay element.
In this regime, a maximum delay variation of up to \SI{3.5}{\nano\second}, corresponding to approximately \SI{10}{\percent} of the nearly constant SW-induced baseline delay, can be achieved. 

\subsection{Power-Dependent RF Characterization}
\label{ssec:A_PowerHandling}

Characterizing the ability to handle high power levels is crucial for RF devices in communication systems on both the transmitter and receiver side.
One of the key figures of merit is the \SI{1}{\decibel} compression point $\PdB$, defined as the input power level at which the output power deviates by \SI{1}{\decibel} from the ideal linear response.
As the presented device is purely passive, the gain $G$ is negative and corresponds to the negative IL.
The $\PdB$ therefore defines the upper boundary of the linear operating regime of the device.
In the present case, it directly reflects the RF power transmission properties of the SW channel, which constitutes the only non-conventional functional element of the overall structure.
The underlying physical effects of the changing SW transmission with increasing input power levels are discussed in more detail in e.g., \cite{Scott_Nonlinear_2004} for a power-dependent change in damping or in \cite{Ustinov_Microwave_2008,Ustinov_Induced_2021} for an additional phase shift caused by non-linear SW excitation.

\begin{figure}[ht!]
    \centering
    \includegraphics[width=0.9\linewidth]{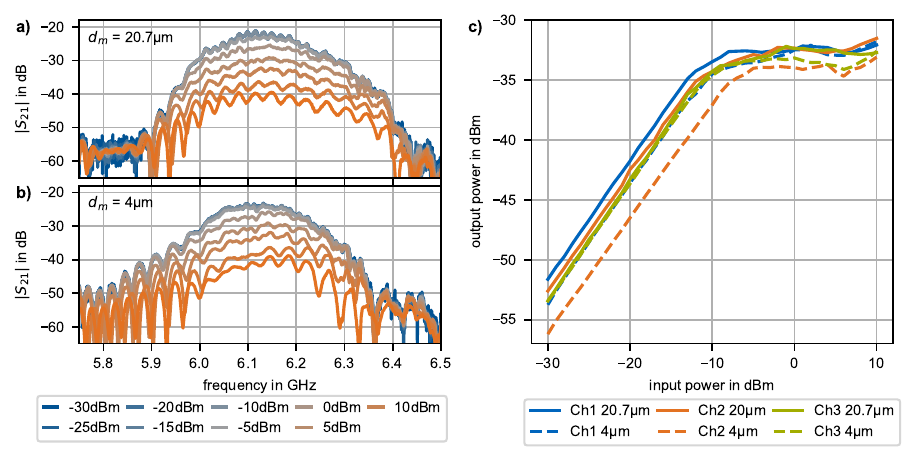}
    \caption{Forward transmission $S_\mrm{21}$ for input powers from \SIrange{-30}{10}{\dBm} at a) $\dm=\SI{20.7}{\micro\meter}$ and b) $\dm=\SI{4}{\micro\meter}$. c) Absolute output vs. input power for all tested channels at the highest and lowest cantilever position at $f_\mrm{c}=\SI{6.1}{\giga\hertz}$. The transmission channels show a typical \SI{1}{\decibel} compression point $\mrm{P}1\mrm{dB}=\SI{-11}{\dBm}$ that marks the end of the regime of linear SW excitation.}
    \label{fig:Hero8_Power_1dB}
\end{figure}

The measured transmission characteristics of the SW channel for input power levels from \SIrange{-30}{10}{\dBm} are shown in Fig.~\ref{fig:Hero8_Power_1dB}a–b for the highest and lowest cantilever positions of the same channel discussed in the main part, respectively.
At the lowest cantilever position $\dm=\SI{4}{\micro\meter}$, the insertion loss is about \SI{2.5}{\decibel} higher than at $\dm=\SI{20.7}{\micro\meter}$.
However, a pronounced increase in IL is observed for input powers above \SI{-10}{\dBm}, at which point the transmitted signal begins to decrease significantly.
The absolute output power $P_\mrm{out}=P_\mrm{in}+G$ at $f_\mrm{c}=\SI{6.1}{\giga\hertz}$ is shown in Fig.~\ref{fig:Hero8_Power_1dB}c for the evaluated channels and for the minimum and maximum cantilever positions.
Except for an outlier at the lowest cantilever position in channel 2, a $\PdB$ of \SI{-11}{\dBm}~$\pm$~\SI{0.5}{\dBm} is obtained across all channels and cantilever positions.
Another relevant metric is the third-order intercept point (IP3), which quantifies the linearity of an RF component by describing the rate at which third-order intermodulation products increase relative to the fundamental signal.
It defines an upper bound of the spurious-free dynamic range in communication systems.
The IP3 was not explicitly evaluated for the magnonic phase shifter, but can be estimated to lie approximately \SIrange{10}{15}{\decibel} above $\PdB$ for most conventional implementations~\cite{Pozar_MicrowaveIP3}.
For higher bias fields, an increase in $\PdB$ is expected, since the threshold for nonlinear spin-wave processes typically scales approximately quadratically with operating frequency.

\subsection{Device Functionality Test at Swept Global Magnetic Field}
\label{ssec:A_FunctionalityTest}

The intended device functionality relies on a two-step process for achieving a phase shift at a given center frequency.
First, the center frequency $f_\mrm{c}$ is set by applying a constant global external bias field $H_\mrm{ext}$.
This center frequency is the operating point of the phase shifter.
Second, at this operating point, the presented phase shifter concept can locally alter the propagating SW characteristics within the magnonic transmission channel.
So, for the intended device functionality being fulfilled, a static global $H_\mrm{ext}$ is needed to set the operating point via a center frequency, and then locally and reversibly changing the phase shift via the stray field $H_\mrm{stray}(x)$ of the micromagnet on the actuated cantilever.
In other words, the phase-shifting capability of the presented device builds on the variation of $H_\mrm{eff}(x)$ within the SW propagation channel, while maintaining a constant $H_\mrm{eff}(x)$ at the input and output transducers' locations.
This behavior translates into the general design goal for every phase shifter: to alter the phase without affecting any other transmission characteristic, such as IL, within the frequency band of interest, which is characterized in Sec.\ref{ssec:A_Variation_fc_IL_BW}.
In contrast to the phase shift evaluation provided in this work, which is related to the actuation of the cantilever, the question arises: how much phase shift can be achieved by applying different global $H_{ext}$ at a targeted center frequency, and how does the phase shift change over a frequency band of interest?

Measurements were performed at both the highest and lowest (not shown) cantilever positions, with the cantilever position kept fixed during the step-wise variation from $\mu_0\Hext$ from \SIrange{148.1}{156.1}{\milli\tesla} in steps of \SI{0.3}{\milli\tesla}.
\num{25} measurements were conducted, \num{7} are shown for visual clarity.
By sweeping $\Hext$, the center frequency $f_\mrm{c}$ shifts according to the in-plane Kittel equation~\cite{Stancil2009}.
Assuming a target center frequency $f_\mrm{c,target}=\SI{6.1}{\giga\hertz}$, the insertion loss at this frequency varies by about \SI{10}{\decibel} over the applied magnetic field range, indicating a shift of the device operating point with $\Hext$.
This behavior shows directly that a change in $H_\mrm{ext}$ has a significant influence on the transmission parameters and is therefore not applicable for a practical use of the device.

\begin{figure}[ht]
    \centering
    \includegraphics[width=0.9\linewidth]{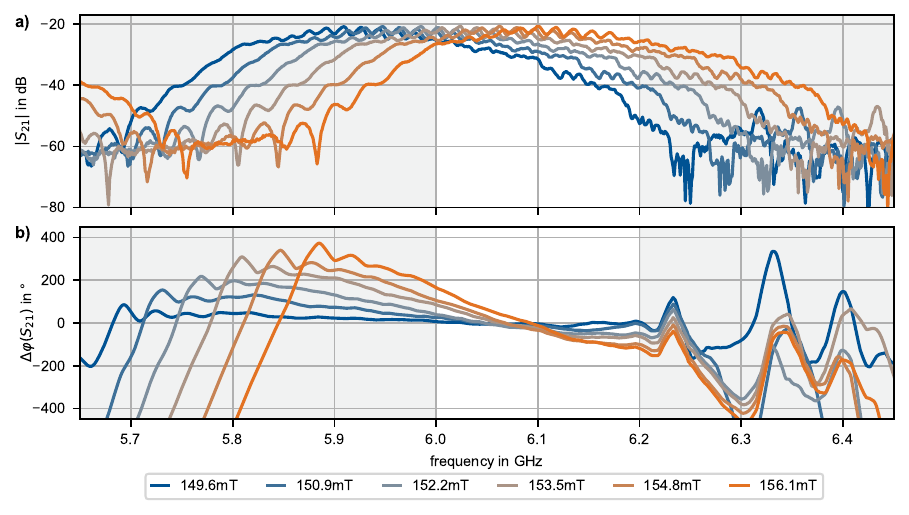}
    \caption{Verification of the phase shift capability by sweeping the global magnetic bias field $\mu_0\Hext$ from \SIrange{148.1}{156}{\milli\tesla} at the highest and lowest (not shown) cantilever position. a) The center frequency shifts with $\Hext$ while at the target center frequency $f_\mrm{c,target}=\SI{6.1}{\giga\hertz}$ the magnitude changes by about \SI{10}{\decibel} in the characterized range. b) The locally changing slope of the phase versus frequency is clearly visible in the phase shift $\Delta\varphi(S_{21})$. A strong frequency dependency is visible in the evaluated phase changes, which makes a SW-related phase shift via the change of the external global bias field $H_\mrm{ext}$ technically unfeasible for broadband RF applications.}
    \label{fig:Hero8_Hsweep_CombiPlot}
\end{figure}

In a next step, we applied the same signal-processing procedure as described in Sec.~\ref{ssec:A_RF_Sparams_Phase} and first performed offset correction on the unwrapped phase at $f_\mrm{c,target}=\SI{6.1}{\giga\hertz}$.
The phase measured at the lowest applied bias field $\mu_0\Hext=\SI{148}{\milli\tesla}$, therefore, is used as the reference signal or phase state and is subtracted from all subsequent measurements at higher $H_\mrm{ext}$, yielding the phase variation shown in Fig.~\ref{fig:Hero8_Hsweep_CombiPlot}b.
This phase variation represents the relative phase change at varying $H_\mrm{ext}$ with respect to $\mu_0\Hext=\SI{148}{\milli\tesla}$, whereas the phase is always evaluated at the fixed targeted center frequency $f_\mrm{c,target}=\SI{6.1}{\giga\hertz}$.
Within a \SI{200}{\mega\hertz} span around $f_\mrm{c,target}$, a phase variation of up to $\pm\SI{160}{\degree}$ is observed at the band edges, while the phase shift near $f_\mrm{c,target}$ remains on the order of a few tens of degrees.
In another picture, the achievable phase shift in a continuous-wave measurement at \SI{6}{\giga\hertz} the tested variation of $H_\mrm{ext}$ would result in a phase shift of about $\pm\SI{160}{\degree}$, while at $f_\mrm{c,target}$ the phase shift is in the order of a few tens of degrees.

For a realistic broadband application of an RF phase shifter, the phase change must be as constant as possible across the frequency band of interest, which cannot be fulfilled when changing the global $H_\mrm{ext}$, but is only possible with the local influence of $H_\mrm{stray}$.
Therefore, a phase shifter that would rely on the change of $H_\mrm{ext}$ would be a highly dispersive and undefined device that even reverses the direction of phase shift, as can be seen from the frequency band edges of interest at \SI{6}{\giga\hertz} and \SI{6.2}{\giga\hertz}.
Combined with the IL change of about \SI{10}{\decibel} by a change of $\mu_0H_\mrm{ext}$ of about \SI{8}{\milli\tesla}, it becomes clear that a global bias field change cannot induce a phase shift that meets the requirements for a technically usable phase shifter operating at a fixed center frequency, shifting the phase almost constant across the bandwidth of interest while leaving other transmission parameters as unchanged as possible. 

\section{Comparison of Analytical Model, Micromagnetic Simulation, and RF Measurements}
\label{sec:A_ModelComparison}

\subsection{Tunability of the Phase Shifter Based on the Lumped-Element Transducer Model}
\label{ssec:A_Tunability_Analytical}

\begin{figure}
    \centering
    \includegraphics[width=0.9\linewidth]{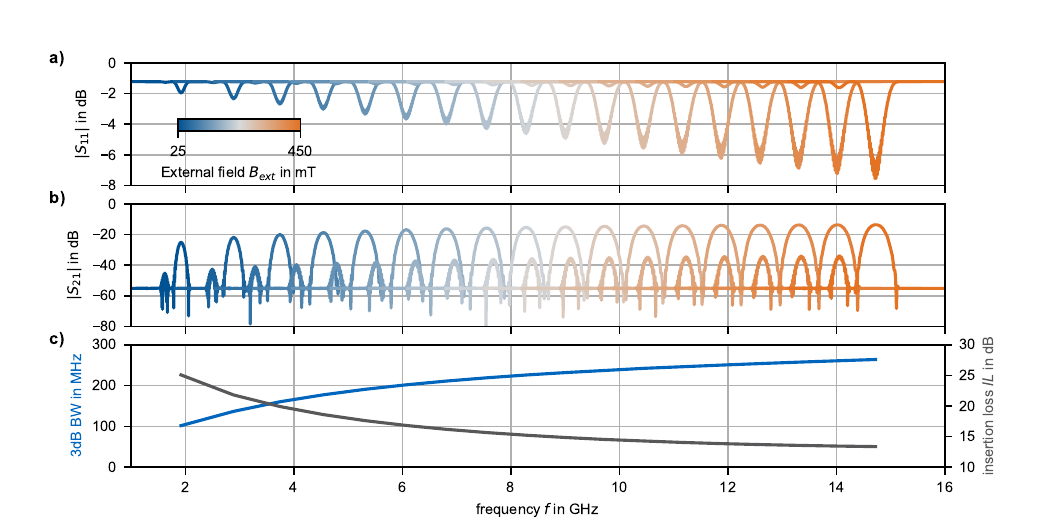}
    \caption{Theoretical evaluation of the tunability of the SW transmission channel of the phase shifter beyond the limited range in the experimental setup. The lumped-element transducer model from~\cite{Kohl_ModellingSW} is used to extract the a) forward reflection $S_{11}$ and b) forward transmission $S_{21}$ of the SW channel including the SW transducers, but without the actuated micromagents. The theoretical evaluation suggests an RL of about \SI{7}{\decibel} and an IL of about \SI{14}{\decibel} around \SI{14.5}{\giga\hertz} or $\mu_0\Hext=\SI{450}{\milli\tesla}$. c) A $\BWthreeDB=\SI{250}{\mega\hertz}$ can be achieved around \SI{14.5}{\giga\hertz} in the model, while experimental data points toward \SI{200}{\mega\hertz} due to phase and amplitude ripples that are not included in the model.}
    \label{fig:Tunability_Model1}
\end{figure}

The lumped-element transducer model enables investigation of the fundamental capabilities of the magnonic phase shifter beyond the experimental limitations imposed by the electromagnet integrated into the wafer prober.
In Fig.~\ref{fig:Tunability_Model1}, we reproduce the prober-based RF characterization of the phase shifter's SW transmission channel (no MEMS) from Sec.~\ref{ssec:Tunability_Comparison} of the main text and extend the analysis up to $\mu_0\Hext=\SI{450}{\milli\tesla}$, corresponding to $f_\mrm{c}\approx\SI{14.5}{\giga\hertz}$.
For this reproduction, a constant electromagnetic crosstalk of approximately \SI{60}{\decibel} is assumed across the full frequency range.
The material and geometrical parameters are chosen as described in Sec.~\ref{sec:Fabrication_Methods} in the main part.
For computational simplicity, the lumped-element model includes modulation of the SW signal by electromagnetic crosstalk but does not account for multi-path propagation effects~\cite{Kohl_ModellingSW,Davidkova2025}, and therefore reproduces only part of the phase and magnitude ripples observed in the experiments.
The model considers only the SW transducers, while the extended CPW feed lines on the magnonic chip are omitted.
In the experiment, ohmic and dielectric losses in these feed lines contribute additional attenuation.
Nevertheless, the agreement between model and measurements allows for a qualitative extrapolation of the device behavior beyond \SI{10}{\giga\hertz}.
From Fig.~\ref{fig:Tunability_Model1}, the return loss decreases with increasing frequency and approaches approximately \SI{8}{\decibel}, indicating more efficient microwave-to-spin-wave conversion.
In other words, the impedance matching between the SW transducer and the effective radiation resistance associated with spin-wave excitation improves at higher frequencies.
A similar trend is observed in Fig.~\ref{fig:Tunability_Model1}b for the transmitted power.
The model indicates an insertion loss up to approximately \SI{14}{\decibel} (also shown on the right axis in Fig.~\ref{fig:Tunability_Model1}c).
While modulation by electromagnetic crosstalk is included, its impact is small for the main SW excitation peak but becomes more visible in the second excitation peak.
As multi-path propagation effects are not included, the simulated transmission spectrum is smoother, leading to an overestimation of $\BWthreeDB$.
As discussed in Sec.~\ref{sec:Operation} of the main text, such effects can be removed using time-domain gating~\cite{Kohl_ModellingSW,Davidkova2025}, although this is not compatible with real-time operation.
Based on the experimental data, an upper limit of approximately \SI{220}{\mega\hertz} for $\BWthreeDB$ around \SI{12}{\giga\hertz} is expected for the present device implementation without post-processing.
In summary, the analysis indicates that the presented magnonic phase shifter remains operable at frequencies well above \SI{10}{\giga\hertz}, with insertion losses below \SI{20}{\decibel} and $\BWthreeDB$ on the order of \SI{200}{\mega\hertz}.

\subsection{Comparison with Hybrid Circuit-Micromagnetic Simulations}
\label{ssec:A_ComparisonMuMax}

\begin{figure}[H]
    \centering
    \includegraphics[width=0.9\linewidth]{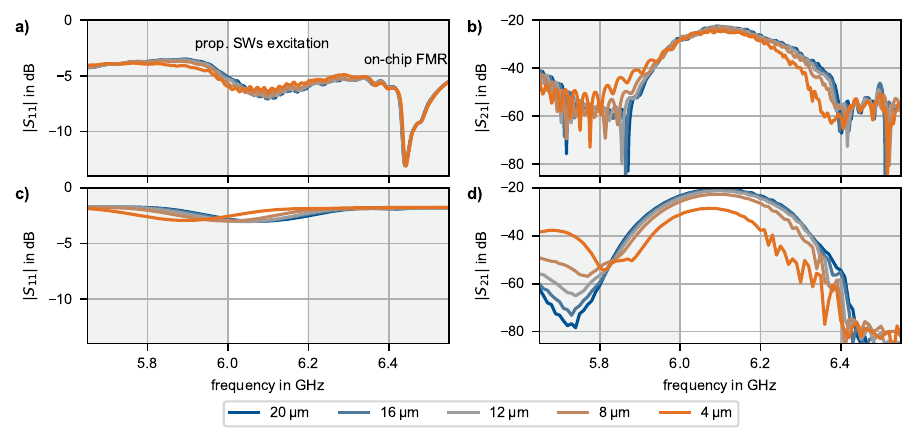}
    \caption{Comparison of the measured a) and simulated c) $|S_{11}|$ as well as the measured b) and d)  $|S_{21}|$ at varying cantilever positions. The simulated S-parameters assume lossless materials for the SW transducers, a perfect short at their ends, and ignore the RF feed lines on the magnonic chip. Within the highlighted \SI{200}{\mega\hertz} span, the simulated and measured $S_{11}$ and $S_{21}$ differ less than about \SI{4}{\decibel} in magnitude, \SI{60}{\mega\hertz} in $\BWthreeDB$, and \SI{25}{\mega\hertz} in $f_\mrm{c}$. At $\dm=\SI{4}{\micro\meter}$, the deviation is most significant due to partial SW reflections at the idealized stray field boundaries in the simulation.}
    \label{fig:Sparameters_SimulationVsMeasurement}
\end{figure}

A side-by-side comparison of the measured forward S-parameters and those obtained from the hybrid circuit–micromagnetic simulation framework is shown in Fig.~\ref{fig:Sparameters_SimulationVsMeasurement}.
In the model, only the SW transducers are included, while the CPW feed lines are neglected.
In the experiment, ohmic and dielectric losses in the CPW feed lines additionally contribute to return losses.
A clear discrepancy is observed between the measured and simulated FMR peak at \SI{6.42}{\giga\hertz}, which originates from feed-line sections fabricated on the YIG film outside the SW transmission channel.
The frequency range of maximum power conversion in the SW system shows only a noticeable deviation at the smallest investigated distance $\dm=\SI{4}{\micro\meter}$, which is attributed to the idealized stray-field profile used in the simulation compared to the measured configuration.
More abrupt spatial variations in the local effective field $\Heff$ can induce additional reflections at the smallest $\dm$.
This trend is consistent with the measured transmission S-parameters in Fig.~\ref{fig:Sparameters_SimulationVsMeasurement}b and the simulated results in Fig.~\ref{fig:Sparameters_SimulationVsMeasurement}d.
The simulated IL of \SI{20.5}{\decibel} is approximately \SI{1.5}{\decibel} lower than the experimental value, as lossy feed-line contributions are not included in the model.
Analogously to Fig.~\ref{fig:RF1_Fcenter_BW_Mag_Combi}a, the simulated center frequency $f_\mrm{c}$ shifts with cantilever position by about \SI{20}{\mega\hertz}, while the $\BWthreeDB$ exceeds \SI{200}{\mega\hertz} at $\dm=\SI{20.7}{\micro\meter}$ and decreases to \SI{175}{\mega\hertz} at $\dm=\SI{6}{\micro\meter}$.
These values are in line with the time-gated data from the prober-based RF characterization.
Only the simulated response at $\dm=\SI{4}{\micro\meter}$ shows a pronounced deviation from the measurements, with $f_\mrm{c}=\SI{6.07}{\giga\hertz}$, $\BWthreeDB=\SI{155}{\mega\hertz}$, and $\mathrm{IL}=\SI{28.5}{\decibel}$.
Compared to the simulated $S_{11}$, the spin-wave back-reflections arising from more abrupt variations in the stray-field profile beneath the soft magnet have a more pronounced impact on the transmission response.
Overall, the simulated and measured results show very good agreement, despite the omission of CPW feed-line effects in the simulation model.

\end{appendices}


\bibliography{sn-bibliography_1}

\end{document}